\pdfoutput=1
\documentclass[sigconf]{acmart}
\AtBeginDocument{%
  \providecommand\BibTeX{{%
    \normalfont B\kern-0.5em{\scshape i\kern-0.25em b}\kern-0.8em\TeX}}}

\copyrightyear{2022}
\acmYear{2022}
\setcopyright{acmcopyright}\acmConference[MM '22]{Proceedings of the 30th ACM International Conference on Multimedia}{October 10--14, 2022}{Lisboa, Portugal}
\acmBooktitle{Proceedings of the 30th ACM International Conference on Multimedia (MM '22), October 10--14, 2022, Lisboa, Portugal}
\acmPrice{15.00}
\acmDOI{10.1145/3503161.3548136}
\acmISBN{978-1-4503-9203-7/22/10}

\usepackage{multirow}
\usepackage{enumitem}
\usepackage{stfloats}
\begin{document}

\title{Disparity-based Stereo Image Compression with Aligned Cross-View Priors}

% \affiliation{%

\author{Yongqi Zhai}
\authornote{Both authors contributed equally to this research.}
\email{zhaiyongqi@stu.pku.edu.cn}
\affiliation{%
  \institution{School of Electronic and Computer Engineering, Peking University}
  \city{Shenzhen}
  \country{China}}

\author{Luyang Tang}
\authornotemark[1]
\email{tly926@stu.pku.edu.cn}
\affiliation{%
  \institution{School of Electronic and Computer Engineering, Peking University}
  \city{Shenzhen}
  \country{China}}

\author{Yi Ma}
\email{mayi@pku.edu.cn}
\affiliation{%
  \institution{School of Electronic and Computer Engineering, Peking University}
  \city{Shenzhen}
  \country{China}}

\author{Rui Peng}
\email{ruipeng@stu.pku.edu.cn}
\affiliation{%
  \institution{School of Electronic and Computer Engineering, Peking University}
  \city{Shenzhen}
  \country{China}}

\author{Ronggang Wang}
\email{rgwang@pkusz.edu.cn}
\authornote{Corresponding author.}
\affiliation{%
  \institution{School of Electronic and Computer Engineering, Peking University}
  \city{Shenzhen}
  \country{China}}

% \affiliation{%

% \affiliation{%

% \affiliation{%

% \affiliation{%

% \affiliation{%

% \affiliation{%

\renewcommand{\shortauthors}{Yongqi Zhai et al.}
\begin{abstract}
With the wide application of stereo images in various fields, the research on stereo image compression (SIC) attracts extensive attention from academia and industry.
The core of SIC is to fully explore the mutual information between the left and right images and reduce redundancy between views as much as possible.
In this paper, we propose DispSIC, an end-to-end trainable deep neural network, in which we jointly train a stereo matching model to assist in the image compression task.
Based on the stereo matching results (i.e. disparity), the right image can be easily warped to the left view, and only the residuals between the left and right views are encoded for the left image.
A three-branch auto-encoder architecture is adopted in DispSIC, which encodes the right image, the disparity map and the residuals respectively.
During training,
the whole network can learn how to adaptively allocate bitrates to these three parts, achieving better rate-distortion performance at the cost of a lower disparity map bitrates.
Moreover, we propose a conditional entropy model with aligned cross-view priors for SIC, which takes the warped latents of the right image as priors to improve the accuracy of the probability estimation for the left image.
Experimental results demonstrate that our proposed method achieves superior performance compared to other existing SIC methods on the KITTI and InStereo2K datasets both quantitatively and qualitatively.
\end{abstract}

\begin{CCSXML}
<ccs2012>
<concept>
<concept_id>10010147.10010371.10010395</concept_id>
<concept_desc>Computing methodologies~Image compression</concept_desc>
<concept_significance>500</concept_significance>
</concept>
</ccs2012>
\end{CCSXML}

\ccsdesc[500]{Computing methodologies~Image compression}

\keywords{stereo image compression, stereo matching, deep learning}


\maketitle

\section{Introduction}
Recently, stereo images have been widely used in various fields, such as 3D movies, virtual reality, autonomous driving and so on. 
These large numbers of high-quality stereo images pose new challenges for data transmission and storage, which require efficient image compression methods to reduce the cost.

\begin{figure}[t]
  \centering
  \includegraphics[width=\linewidth, trim = {220 117 230 85}, clip]{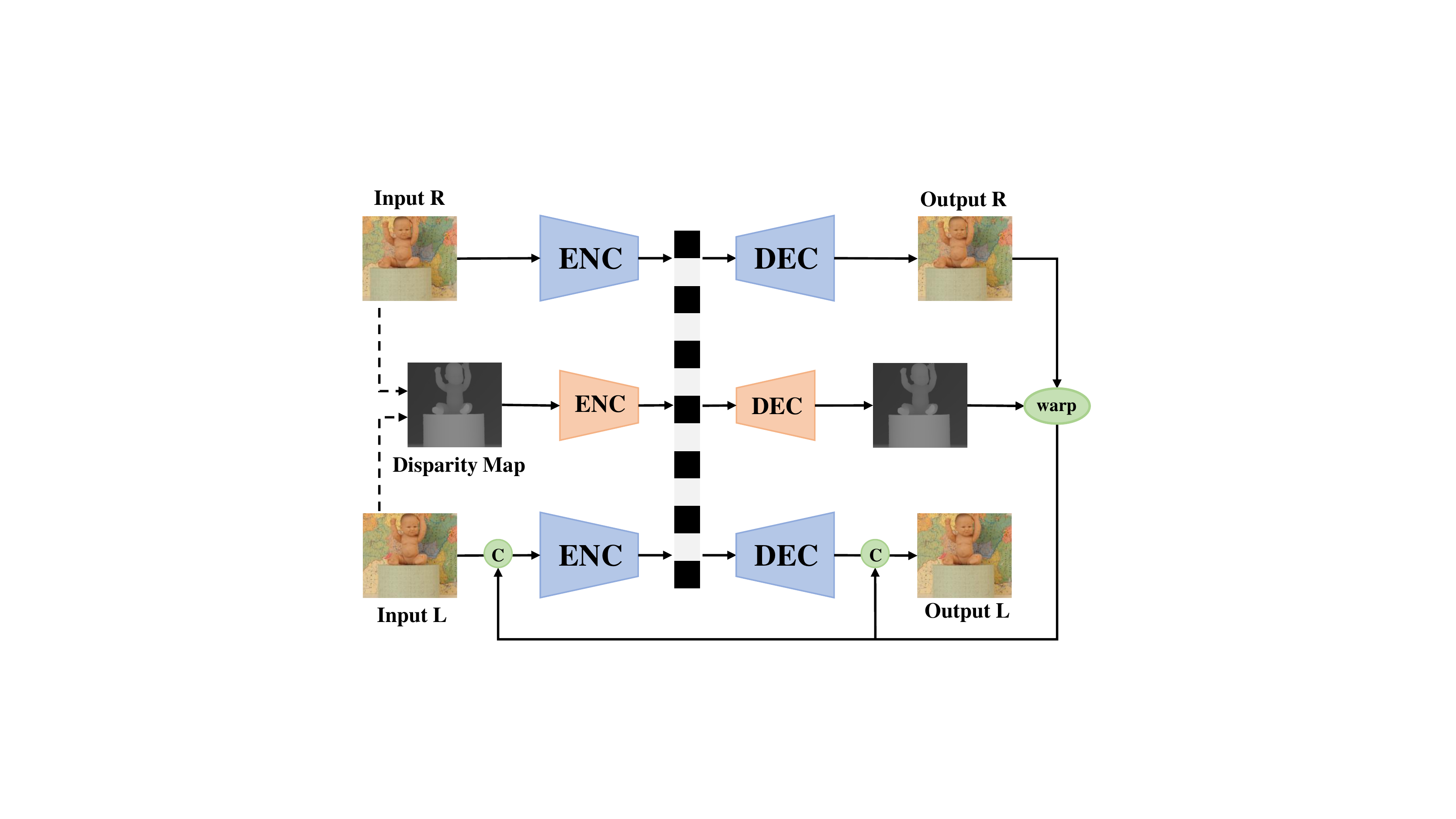}
  \caption{Brief framework of our proposed DispSIC. A three-branches is adopted, which encodes the input right image, the disparity map and the residuals of the left image respectively.}
  \label{brief_framework}
\end{figure}

Image compression is a fundamental and vital research topic in the field of multimedia. After decades of research, a lot of traditional image compression standards have been proposed, such as JPEG \cite{1992The}, JPEG2000 \cite{2000The}, BPG \cite{bpgurl} and VVC-intra \cite{VVC}. Typically, they follow the pipeline of three hand-crafted modules: transformation, quantization and entropy coding. Recently, benefiting from the end-to-end optimization,  deep neural network (DNN) based image compression methods have achieved promising results \cite{balle2017end-to-end,balle2018variational,minnen2018joint,lee2019context-adaptive,cheng2020image,hu2020coarse,he2021checkerboard,xie2021enhanced,zhu2021transformer, qian2022entroformer}. 

A stereo image pair consists of left and right images, which are captured by a stereo camera from both views at the same time.
For stereo images, a simple idea is to compress each image separately using a single image compression method.
However, due to the large overlapping fields of view between the left and right cameras, quite a lot of similar information inherently exists between stereo images.
This idea ignores the high correlation between views and wastes a lot of bitrates, which is rather unwise. 
Therefore, it is necessary to design a compression method that can make full use of the inner relationship and minimize the redundancy between views as much as possible.

Stereo image compression (SIC) compresses the left and right images jointly, aiming to achieve high compression ratios for both two images. At present, there have been many studies focusing on SIC.
In many traditional SIC methods \cite{merkle2006efficient, lukacs1986predictive}, the disparity map is often compressed additionally as important side information to reduce the bitrates.
Given a reference view, they use the disparity compensated prediction to encode other views.
However, traditional methods minimize the rate-distortion loss based on hand-crafted features, which limits their compression efficiency.
Recently, deep learning-based SIC methods have achieved better performance than traditional ones.
Liu et al. \cite{liu2019dsic} proposed the first deep stereo image compression network, namely DSIC, which densely warps the features from the left image to the right at all levels in the encoder and decoder to exploit the content redundancy between the stereo pair.
When training and testing, DSIC requires to construct cost volume and aggregate the cost multiple times, which has high computation complexity.
After that, Deng et al. \cite{deng2021deep} propose an efficient stereo image compression based on homography transformation (HESIC), which reduces the computation complexity and achieves better compression performance.
In HESIC, the left image is spatially transformed by a $3\times3$ homography matrix and the residuals between the right and the transformed left image are encoded.
However, the homography matrix is used to describe the relationship between two images of the same planar surface in space. Most stereo image pairs have difficulty meeting this condition, which will make the residuals in HESIC very large.
Compared with the homography matrix, the disparity map establishes a more accurate pixel-to-pixel correspondence between stereo images, which can significantly reduce the residual information. Moreover, as a simple 1D data flow, the disparity map is easily compressed. 

In this paper, we propose a disparity-based deep neural network for stereo image compression, namely DispSIC, which combines the disparity compensated prediction in traditional methods with a deep learning framework.
As Fig.~\ref{brief_framework} shows, we replace the homography matrix in HESIC \cite{deng2021deep} with the disparity map, which is output by a stereo matching model.
Unlike the traditional SIC methods using disparity compensated prediction, we train the stereo matching model and the compression model jointly with the rate-distortion loss. 
By end-to-end optimization, the stereo matching model learns how to provide more accurate disparity compensated prediction, and the compression model learns how to trade off the transmission cost and gain of the disparity map.
The two models work toward the same objective i.e. reconstructing higher quality images with fewer bitrates.

In conclusion, the main contributions of this paper are as follows:
\begin{itemize}[leftmargin=*]
\item We propose a novel disparity-based deep neural network for stereo image compression, namely DispSIC, which uses the disparity map to explicitly represent the relationship between the left and right images.
\item We design a conditional entropy model with aligned cross-view priors, which can model the probability of the left image more accurately.
\item The experimental results show that our proposed method significantly outperforms the state-of-art deep stereo image compression methods, and saves around 25\% bitrates compared to the latest method HESIC \cite{deng2021deep} with similar image quality.
\end{itemize}

\begin{figure*}[t]
\begin{center}
\includegraphics[width=0.92\linewidth, trim={10, 79, 10, 12.5}, clip]{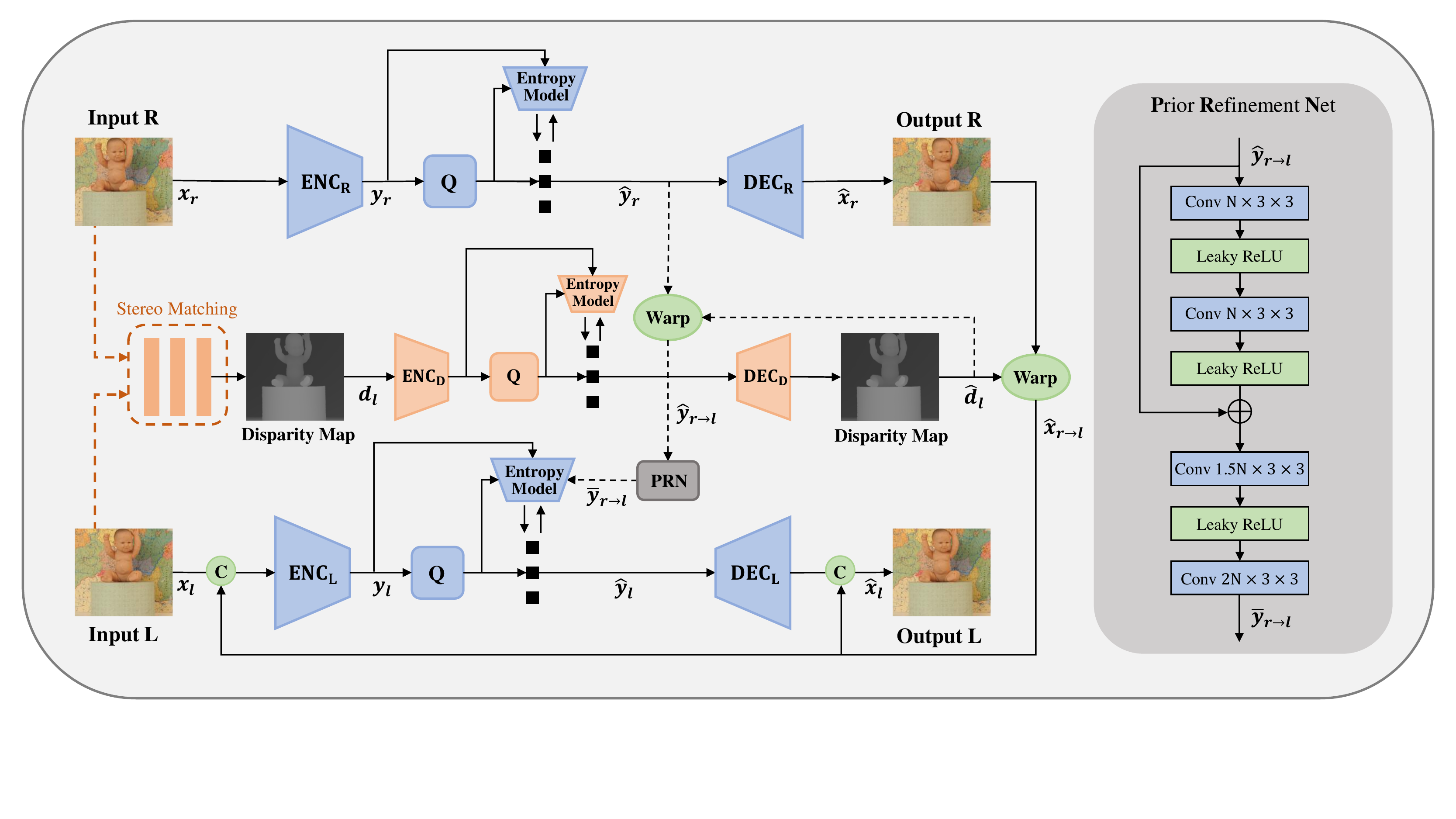}
\end{center}
  \caption{The overall network architecture of our proposed method (DispSIC). We compress the left and right images jointly, and use the disparity map to explicitly represent the pixel-wise correlation between views to save bitrates. C represents the concatenation operation, Q represents quantization.}
\label{overall_framework}
\end{figure*}

\section{Related Work}
\subsection{Single Image Compression} 
Existing traditional image compression standards include JPEG \cite{1992The}, JPEG2000 \cite{2000The}, BPG \cite{bpgurl} and VVC-intra \cite{VVC}. The pipeline of these methods typically consists of transformation, quantization, and entropy coding. Although these traditional methods can achieve high compression efficiency, they heavily rely on prior knowledge to design hand-crafted modules. 
In recent years, the DNN-based image compression methods \cite{balle2017end-to-end,balle2018variational,minnen2018joint,lee2019context-adaptive,cheng2020image,hu2020coarse,he2021checkerboard,xie2021enhanced,zhu2021transformer, qian2022entroformer} have achieved great success with impressive performance. Unlike traditional methods, they jointly optimize all modules in an end-to-end manner. For the network architecture, some early works used the recurrent neural networks (RNNs) to achieve progressive compression \cite{toderici2015variable,su2020scalable}. Then, most works used the convolutional neural networks (CNNs) to build an auto-encoder style network. Most recently, there have been some works employing the invertible neural networks (INNs) \cite{xie2021enhanced} and transformer \cite{zhu2021transformer, qian2022entroformer} architecture. Besides, some works \cite{theis2017lossy, balle2017end-to-end,agustsson2017soft} were proposed to solve the problem of non-differential quantization and rate estimation. Other works such as generalized divisive normalization (GDN) \cite{balle2017end-to-end}, attention mechanism \cite{liu2019non,cheng2020image},  residual blocks \cite{cheng2020image}, adaptive feature extraction models \cite{ma2021afec}, adversarial training \cite{rippel2017real,agustsson2019generative,mentzer2020high}, and importance map \cite{mentzer2018conditional,zhang2021attention} were focused on improving the image compression performance. Meanwhile, variable-rate model \cite{choi2019variable,cui2021asymmetric}, scalable compression \cite{jia2019layered,mei2021learning} were introduced to meet the practical application requirements.

\subsection{Stereo Matching} 
Stereo matching is the process of finding pixels corresponding to the same 3D point in the scene, consisting of four steps: matching cost computation, cost aggregation, disparity optimization and post-processing \cite{scharstein2002taxonomy}.
As we know, the core of SIC is to explore the inner correlation between the left and right images, in which stereo matching techniques play an important role.

Traditional stereo matching methods can be roughly divided into three categories: local methods \cite{birchfield1999depth, ryan1980prediction}, global methods \cite{kolmogorov2001computing, klaus2006segment} and semi-global methods \cite{hirschmuller2007stereo}.
Local methods compute the matching cost on a block-by-block or pixel-by-pixel basis, which run fast but have low-quality results.
Global methods first obtain an initial disparity map through simple matching, and then iteratively update the disparity map by optimizing a pre-defined global energy function. These methods can get high-quality results but are time-consuming.
Semi-global matching (SGM) methods perform line optimization along multiple directions, which successfully combine local and global stereo methods and achieve a favorable trade-off between computing time and quality of the results.
Although traditional methods have made great progress, they do not work well in regions of low texture, repeating patterns and thin structures.
Nowadays, with the development of deep learning, CNN-based stereo matching methods \cite{mayer2016large, kendall2017end, chang2018pyramid} achieve promising performance even in the above challenging regions. 
In terms of the network structure, existing end-to-end stereo matching methods include 2D \cite{liang2018learning, yang2018segstereo, song2020edgestereo} and 3D \cite{yu2018deep, guo2019group} convolution-based neural networks, differing in the way of cost aggregation. In comparison, most of the 3D-CNN-based methods perform better but are more time-consuming.
In terms of training methods, most studies \cite{luo2016efficient, zbontar2016stereo} use the ground-truth disparity as supervision to train the network. However, given that the ground truth is not always available, some studies \cite{zhong2017self, pilzer2019progressive} warp the left image to the right view based on the estimated disparity map and compute the reconstruction error as supervision.

\subsection{Stereo Image Compression} 
% 补充传统SIC方法 %
Traditional SIC methods \cite{flierl2007multiview, merkle2006efficient, martinian2006view, kitahara2006multi,ellinas2004stereo, bezzine2018sparse, kadaikar2018joint} usually compress stereo images in the way of video compression, which regards the inter-view similarity between the stereo camera views as the temporal similarity between successive frames in the video.
Given a reference view, many works \cite{merkle2006efficient, lukacs1986predictive} use the disparity compensated prediction to encode each view, which is similar to the motion-compensated prediction in the single-view video.
These methods rely on hand-crafted features and design modules individually, which greatly limits their compression performance.

Recently, several DNN-based methods for SIC have been proposed, which outperform the traditional methods.
Liu \cite{liu2019dsic} proposed the first DNN-based stereo image compression framework (DSIC), which uses parametric skip functions to feed the disparity-warped features at all levels from the left encoder/decoder to the right one.
Due to the dense warp scheme, it has high computation complexity.
After that, Deng et al . \cite{deng2021deep} use a homography transformation to replace the dense warp, which reduces the computation complexity and achieves state-of-art performance.
In HESIC \cite{deng2021deep}, a $3\times3$ homography matrix is used to spatially transform the left image to the right view. And only the residuals between the original right image and the transformed left image are encoded to reconstruct the right image.
However, only when two images are of the same planar surface in space, the homography matrix can provide an accurate match. So in most scenarios, the homography transformation will lead to large residuals in HESIC \cite{deng2021deep}.

\begin{figure*}[t]
  \centering
  \includegraphics[width=0.93\linewidth, trim = {0 290 0 0}, clip]{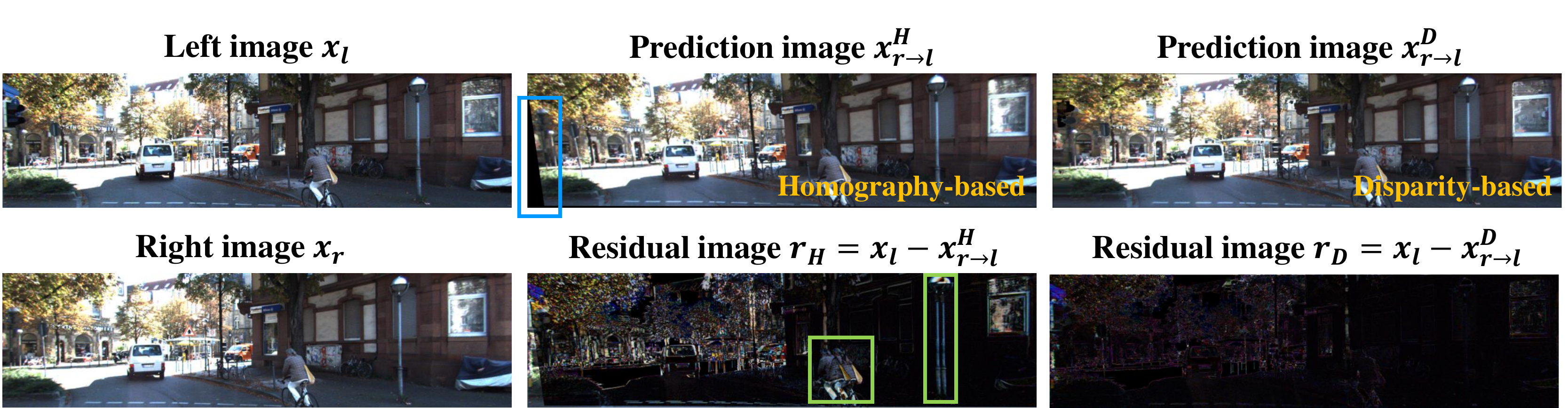}
  \caption{Visual comparisons of images generated by disparity-based warping and homography-based transformation. The blue box shows that the homography-based prediction image $x_{r{\rightarrow}l}^H$ misses some border pixels. The green boxes show that the "Bicyclist" and "Pole" areas in $x_{r{\rightarrow}l}^H$ misalign with the original texture.}
  \label{homograpy_disparity}
\end{figure*}

\section{Methods}

\subsection{Framework}

Fig.~\ref{overall_framework} shows the overall framework of our proposed DispSIC network. 
In general, the whole network adopts a three-branch structure, which is used to encode the right image, the disparity map and the residuals of the left image respectively.

Firstly, taking the original left and right images (denoted as $x_l$ and $x_r$) as input, we use a stereo matching model to estimate the disparity map (denoted as $d_l$).
Then, $x_r$ and $d_l$ are compressed via two independent auto-encoders \cite{minnen2018joint}. Based on the decoded disparity map $\hat{d}_l$, the decoded right image $\hat{x}_r$ can be easily warped to the left view, obtaining the predicted left image $\hat{x}_{r{\rightarrow}l}$. 
After that, we concatenate the original left image $x_l$ and the predicted $\hat{x}_{r{\rightarrow}l}$ as the input to the third auto-encoder \cite{minnen2018joint}, which learns to compress the residuals.
Taking into account the correlation between $x_l$ and $x_r$, we take the warped quantized latent of $x_r$ as priors for the entropy model of the latent of $x_l$.
At the decoder side, the decoded residuals are concatenated with the predicted 
$\hat{x}_{r{\rightarrow}l}$ to reconstruct the final left image $\hat{x}_l$.

\begin{figure}
  \centering
  \includegraphics[width=0.95\linewidth, trim = {150 17 150 0}, clip]{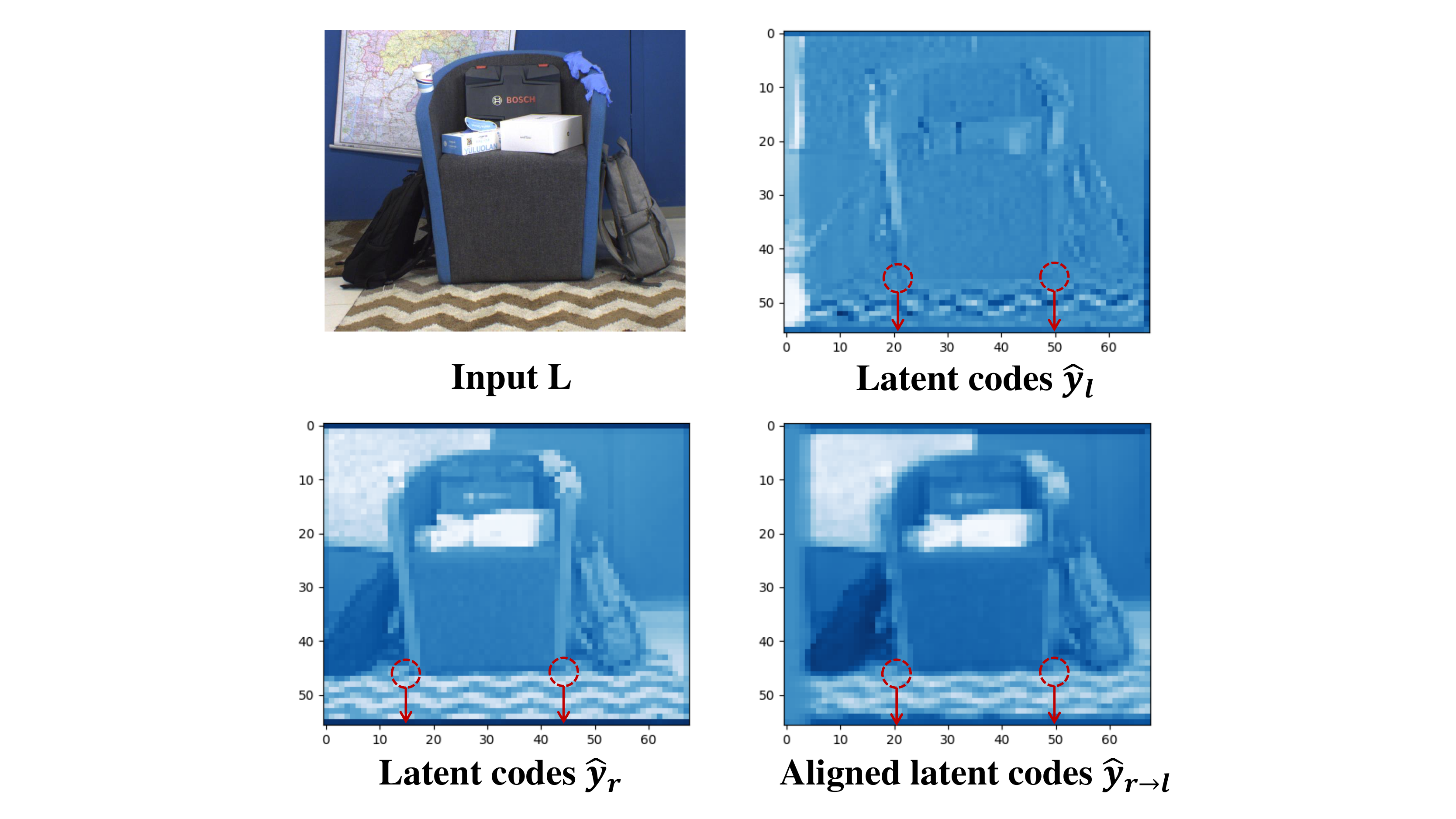}
  \caption{Visualization of latent codes of the left and right images. The tick marks represent the coordinates of the pixel.}
  \label{entropy_image}
\end{figure}
\subsection{Disparity-based Prediction} \label{disparity-based prediciton}
To fully explore the inner relationship between stereo images, we need to match the images or features first.
DSIC \cite{liu2019dsic} constructs cost volume at each feature level to match the features between stereo images, which works but has high computation complexity.
HESIC \cite{deng2021deep} uses a $3\times3$ homography matrix to roughly match the stereo images, which is simple but inaccurate.
As Fig.~\ref{homograpy_disparity} shows, the image generated by the homography transformation has many problems, such as border pixels missing and misaligned textures with the original image, which greatly increase the residual information.
In contrast, the disparity refers to the difference in coordinates of similar features within two stereo images, which can reflect the pixel-level correlation between the left and right images. As a 1D data flow, it is lightweight and can be easily compressed.
Fig.~\ref{homograpy_disparity} shows that the image generated by disparity-based warping is of better quality and has less residual information.
Therefore, we propose to use the disparity map to assist with stereo image compression.

For convenience, we adopt an existing stereo matching model \cite{xu2020aanet}.
Taking the left image $x_l$ and the right image $x_r$ as input, it can generate the disparity map $d_l$ aligned with $x_l$.
% Firstly, a shared feature extractor is used to extract the downsampled feature pyramid of $x_l$ and $x_r$.
% Then, a cost volume is constructed by correlating left and right image features at each scale.
% After cost aggregation and disparity regression, the disparity map $d_l$ is acquired.
On the decoder side, based on the decoded disparity map $\hat{d}_l$, following \cite{jaderberg2015spatial} we can obtain the left predicted image 
$\hat{x}_{r{\rightarrow}l}$ by using a bilinear sampler to sample the decoded right image $\hat{x}_r$ with backward mapping.
The whole process is fully differentiable and can integrate into our end-to-end trainable compression network.

\subsection{Aligned Cross-View Priors} \label{entropy model}
Given the latent $\hat{y}$, the entropy model $p_{\hat{y}}$ is used to fit the marginal distribution $m({\hat{y}})$. The smallest average code length (bitrates) is decided by the matching degree of $p_{\hat{y}}$ and $m({\hat{y}})$, given by the Shannon cross entropy between the two distributions:
\begin{equation}
   R = \mathbb{E}_{{\hat{y}\sim {m}}}[-{\log_2}p_{\hat{y}}({\hat{y}})].
\end{equation}

For the quantized hyper latent $\hat{z}_l$, $\hat{z}_r$ and $\hat{z}_d$, we use a non-parametric, fully factorized density model \cite{balle2017end-to-end}. For the right quantized latent $\hat{y}_r$ and disparity map quantized latent $\hat{y}_d$, we use an autoregressive context model with a mean and scale hyperprior \cite{minnen2018joint}, which can be formulated as:
\begin{equation}
    \begin{split}
  p_{\hat{y}_r |\hat{z}_r}(\hat{y}_r |\hat{z}_r) \sim   \mathcal{N}(\mu_r,{\sigma}^2_r), \\
    p_{\hat{y}_d |\hat{z}_d}(\hat{y}_d |\hat{z}_d) \sim   \mathcal{N}(\mu_d,{\sigma}^2_d),
  \label{eq:prob_y_l}
  \end{split}
\end{equation}
where ${\mu}$, ${\sigma}$ are the estimated mean and scale parameters.

Due to the large overlapping fields of view between the left and right cameras, 
strong correlation exists between the left quantized latent $\hat{y}_l$ and right quantized $\hat{y}_r$.
We visualize the features of the left and right images in Figure~\ref{entropy_image}. 
It is obvious that $\hat{y}_l$ and $\hat{y}_r$ have structural similarities,
so the mutual information ${I(\hat{y}_l, \hat{y}_r)}$ is positive.
We denotes the Shannon entropy of $\hat{y}_l$ as ${H(\hat{y}_l)}$,  and then the conditional entropy ${H(\hat{y}_l|\hat{y}_r})$ can be obtained: 
\begin{equation}
   H(\hat{y}_l|\hat{y}_r) = H(\hat{y}_l) - I(\hat{y}_l, \hat{y}_r) < H(\hat{y}_l).
\end{equation}

Since $\hat{y}_r$ is transmitted before $\hat{y}_l$, $\hat{y}_r$ can serve as the cross-view priors to model the distribution $p_{\hat{y}_l}$.

Moreover, considering that the disparity also exists between $\hat{y}_l$ and $\hat{y}_r$, e.g. Figure~\ref{entropy_image} shows that the bag features in $\hat{y}_l$ have horizontal position offsets compared to $\hat{y}_r$, 
directly feeding $\hat{y}_r$ as priors into the entropy estimation network of $\hat{y}_l$ like \cite{liu2019dsic,deng2021deep} is not accurate enough.
So we warp the $\hat{y}_r$ based on the decoded disparity map $\hat{d}_l$ to get aligned latent codes $\hat{y}_{r\rightarrow{l}}$. 
% The warping operation may introduce some spatial discontinuity, we design a prior refinement net (PRN) $f_{pr}(\cdot)$ to refine the aligned cross-view priors $\overline{y}_{r\rightarrow{l}} = f_{prn}(\hat{y}_{r\rightarrow{l}})$. 
Besides, we design a prior refinement net (PRN) $f_{pr}(\cdot)$ to refine $\hat{y}_{r\rightarrow{l}}$, removing the spatial discontinuity introduced by the warping operation.
The generation process of the final aligned cross-view priors can be formulated by:
\begin{equation}
    \overline{y}_{r\rightarrow{l}} = f_{prn}(warp(\hat{y}_r,\hat{d}_l)).
\end{equation}

\begin{figure}
  \centering
  \includegraphics[width=0.93\linewidth, trim = {220 155 275 135}, clip]{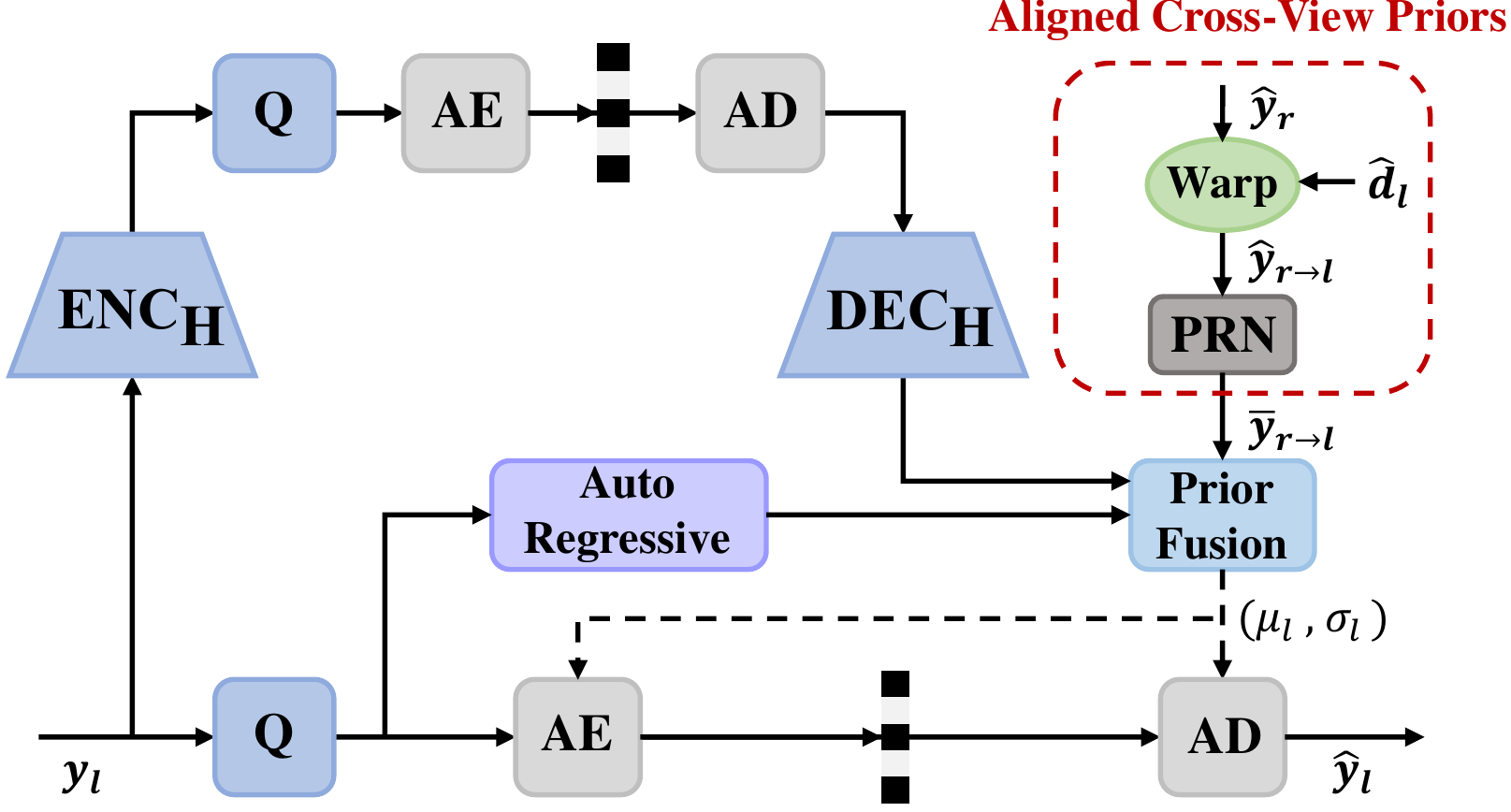}
  \caption{ Our conditional entropy model used to encode the quantized latent $\hat{y}_l$. ${ENC}_H$ and ${DEC}_H$ represent the  hyperprior encoder and decoder. AE and AD are the arithmetic encoder and arithmetic decoder.}
  \label{cross-view}
\end{figure}
% The framework of our conditional entropy model is illustrated in Fig. ~\ref{cross-view}. 
Figure~\ref{cross-view} illustrates the structure of our conditional entropy model.
We use the hyperprior model \cite{minnen2018joint} and the auto regressive network \cite{minnen2018joint} to learn the hierarchical and spatial priors, respectively.
Then we fuse these two priors with the aligned cross-view priors.
We model the latent $\hat{y}_l$ as a conditional Gaussian distribution:
\begin{equation}
  p_{\hat{y}_l |\hat{z}_l,\overline{y}_{r\rightarrow{l}}}(\hat{y}_l |\hat{z}_l,\overline{y}_{r\rightarrow{l}}) \sim   \mathcal{N}(\mu_l,{\sigma}^2_l),
  \label{eq:prob_y_r}
\end{equation}
where $\hat{z}_l$ represents the left quantized hyper latents, ${\mu_l}$ and ${\sigma_l}$ are the estimated mean and scale parameters of left quantized latents $\hat{y}_l$. The rate of the left image ${R}_l$, the right image ${R}_r$ and the disparity map ${R}_d$ can be written as:
\begin{equation}
  \begin{split}
  R_{{l}} &= \mathbb{E}[-{\log_2}p_{\hat{y}_l |\hat{z}_l,\overline{y}_{r\rightarrow{l}}}(\hat{y}_l |\hat{z}_l,\overline{y}_{r\rightarrow{l}})] 
  + \mathbb{E}[-{\log_2}p_{\hat{z}_l}({\hat{z}_l})],    \\
 R_{{r}} &= \mathbb{E}[-{\log_2}p_{\hat{y}_r |\hat{z}_r}(\hat{y}_r |\hat{z}_r)] + \mathbb{E}[-{\log_2}p_{\hat{z}_r}({\hat{z}_r})],  \\
  R_{{d}} &= \mathbb{E}[-{\log_2}p_{\hat{y}_d |\hat{z}_d}(\hat{y}_d |\hat{z}_d)] + \mathbb{E}[-{\log_2}p_{\hat{z}_d}({\hat{z}_d})].
  \end{split}
\end{equation}

\begin{figure*}[t]
\begin{center}
\includegraphics[width=0.96\linewidth, trim={0, 15, 0, 21}, clip]{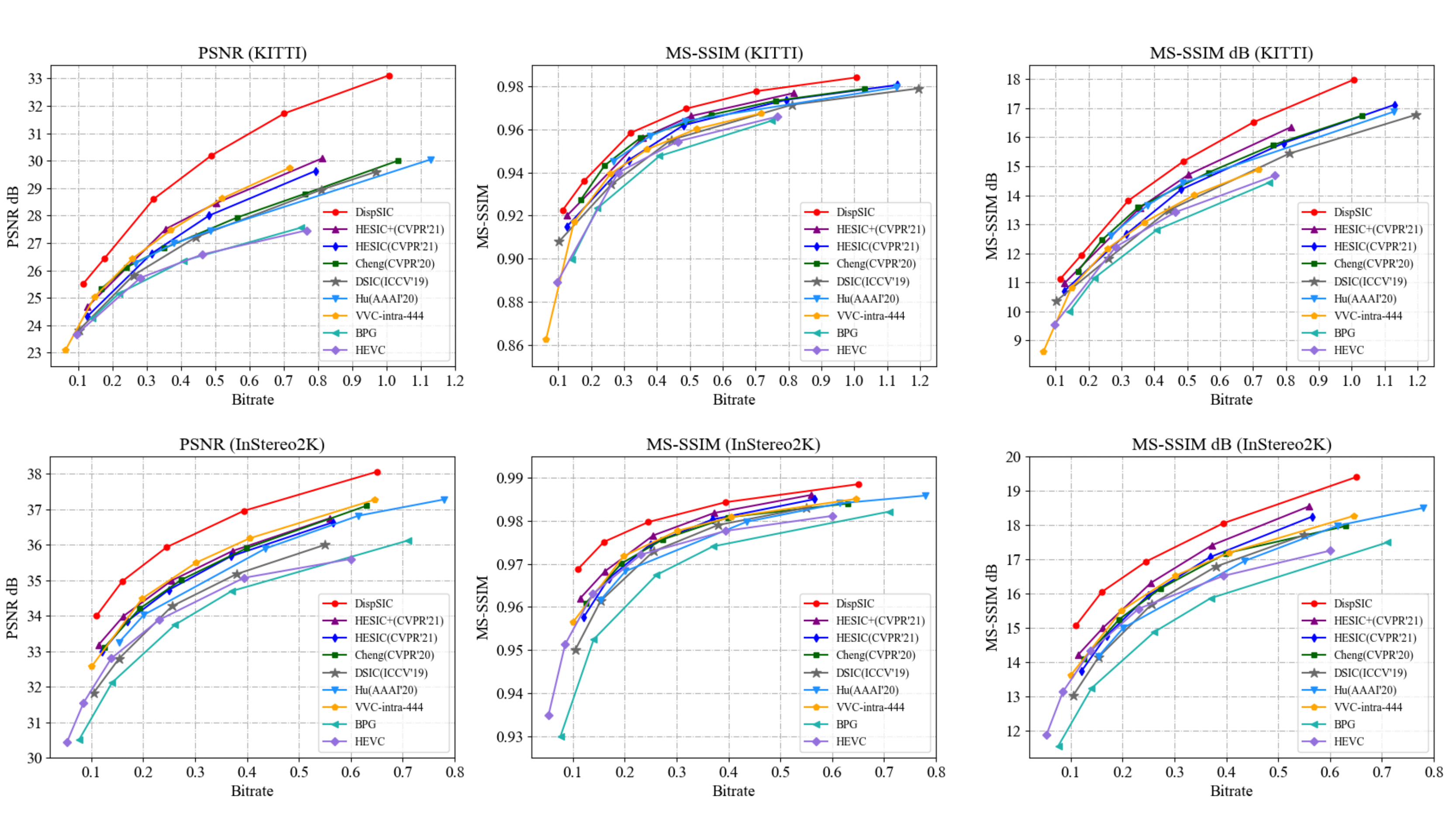}
\end{center}
  \caption{Rate-distortion performance comparison of different methods on the KITTI and InStereo2K datasets.}
\label{psnr_ssim}
\end{figure*}

\subsection{Training Strategy}
Image compression aims to reconstruct the image of best quality with the least bitrates, which is a Lagrangian multiplier-based rate-distortion optimization problem. The loss function is formulated as:
\begin{align}
\label{equation_loss}
 &  \mathcal{L} =R + \lambda D,
%  & \qquad with \quad w(i) = i Mod 8 
\end{align}
where $R$ denotes the bitrates, $D$ denotes the distortion and $\lambda$ controls the rate-distortion trade-off. 

To stabilize the training process, we use the pre-trained stereo matching model \cite{xu2020aanet} as the initialization of disparity estimation. 
In addition, we introduce an auxiliary loss $D(x_l, \hat{x}_{r{\rightarrow}l})$ at the early stage of training to instruct the whole network to utilize the disparity compensated prediction. The initial loss function $\mathcal{L}_{init}$ is defined as follows:
\begin{align}
\label{equation_loss}
 &  \mathcal{L}_{init} =R_l+R_r+R_d+\lambda (D(x_l, \hat{x}_l) + D(x_r, \hat{x}_r) + \beta D(x_l, \hat{x}_{r{\rightarrow}l})),
%  & \qquad with \quad w(i) = i Mod 8 
\end{align}
where the hyperparameter $\beta$ is set to $0.2$. When the training loss $\mathcal{L}_{init}$ is stable, the auxiliary loss $D(x_l, \hat{x}_{r{\rightarrow}l})$ is removed and $\beta$ is set to $0$. The total rate-distortion loss $\mathcal{L}_{final}$ can be formulated as:
\begin{align}
\label{equation_loss}
 &  \mathcal{L}_{final} =R_l+R_r+R_d+\lambda (D(x_l, \hat{x}_l) + D(x_r, \hat{x}_r)).
%  & \qquad with \quad w(i) = i Mod 8 
\end{align}

\begin{figure*}[t]
\begin{center}
\includegraphics[width=0.922\linewidth, trim={10, 134, 15, 130}, clip]{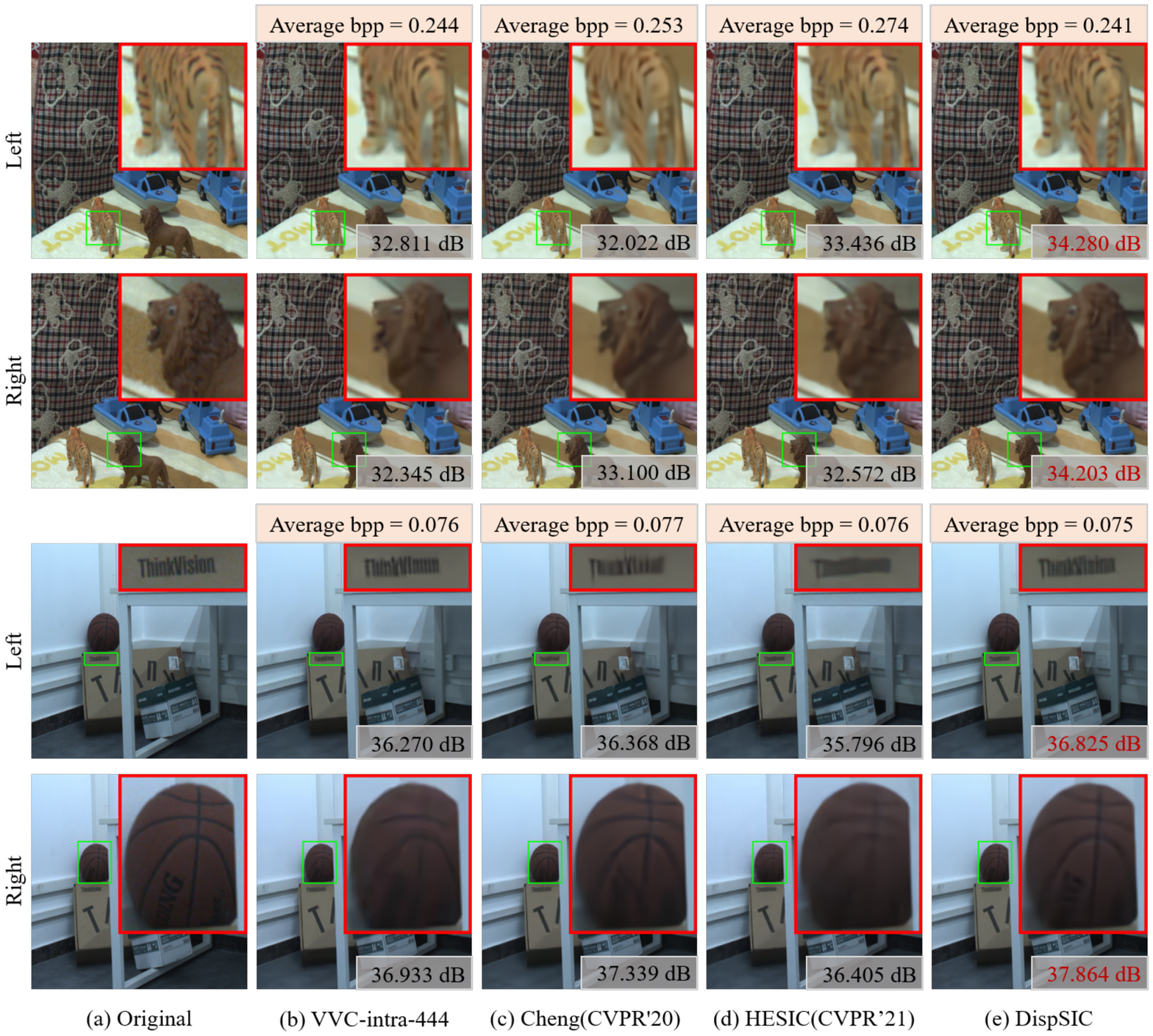}
\end{center}
  \caption{Qualitative comparison.
  It is worth mentioning that these images are not cherry-picked, they are the same as images in HESIC\cite{deng2021deep}.}
\label{qualitative_results}
\end{figure*}

\section{Experiments}
\subsection{Experimental Setup}
\textbf{Datasets.} 
We evaluate the compression performance of our proposed DispSIC on two public stereo image datasets: 
% \begin{enumerate}
\begin{enumerate}[leftmargin=*]
\item KITTI \cite{geiger2013vision}: A dataset of outdoor scenes. The stereo images in it are with far views. We use 1,950, 50 and 50 stereo image pairs which are randomly selected by HESIC \cite{deng2021deep} for training, validation and testing respectively.
\item InStereo2K \cite{bao2020instereo2k}: A dataset of indoor scenes. It consists of 2050 stereo image pairs. The stereo images in it are with close-views. Likewise, 1,950 pairs are used for training and 50 pairs for validation. The remaining 50 pairs are used for testing.
\end{enumerate}
During training, the images are randomly cropped to $256\times512$ patches. By using these two datasets of different scenarios, we can evaluate our method comprehensively.

\textbf{Implementation details.}
We implement our framework on the CompressAI PyTorch library \cite{begaint2020compressai}. We optimize our network using Adam optimizer \cite{kingma2014adam} with a batch size of $4$. We first train our model on KITTI, and then the model is fine-tuned to InStereo2K. All models are trained on a single V100 GPU for 1600 epochs. The learning rate is $1\times10^{-4}$ for 1200 epochs and reduced to $1\times10^{-5}$ for the last iterations. 
We  train our network with the loss $\mathcal{L}_{init}$ in the first $800$ epochs, and then with the loss $\mathcal{L}_{final}$ in the remaining epochs. 
We use mean square error (MSE) as our quality metric, and train 5 models for both datasets. The $\lambda$ is chosen from set $\left\{ 0.001, 0.002, 0.005, 0.01, 0.02\right\}$.

\begin{table}
  \caption{ BD-PSNR and BD-rate comparisons, with the best results in red.}
  \label{bdbr}
  \begin{tabular}{|c|cc|}
    \hline
    \multicolumn{3}{|c|}{\textbf{KITTI dataset}}\\
    \hline
    Methods & BD-PSNR(dB)$\uparrow$ & BD-rate(\%)$\downarrow$\\
    \hline
    BPG & -0.533 & 22.624 \\
    HEVC/H.265 & -0.412 & 15.665 \\
    Hu (AAAI'20) & 0.004 & 5.510 \\
    HESIC (CVPR'21) & 0.357 & -13.296 \\
    Cheng (CVPR'20) & 0.234 & -7.723 \\
    VVC-intra-444 & 0.691 & -23.113 \\
    HESIC+ (CVPR'21) & 0.862 & -27.069 \\
    DispSIC & \textcolor{red}{2.308} & \textcolor{red}{-51.719} \\
    \hline
    \multicolumn{3}{|c|}{\textbf{InStereo2K dataset}}\\
    \hline
    Methods & BD-PSNR(dB)$\uparrow$ & BD-rate(\%)$\downarrow$\\
    \hline
    BPG & -0.478 & 20.683 \\
    HEVC/H.265 & -0.043 & 0.677 \\
    Hu (AAAI'20) & 0.420 & -15.108 \\
    HESIC (CVPR'21) & 0.625 & -22.19 \\
    Cheng (CVPR'20) & 0.681 & -23.666 \\
    VVC-intra-444 & 0.903 & -30.171 \\
    HESIC+ (CVPR'21) & 0.838 & -28.920 \\
    DispSIC & \textcolor{red}{1.847} & \textcolor{red}{-53.544}\\
    \hline
\end{tabular}
\end{table}

\begin{table}
\begin{center}
\caption{Computational complexity comparison.}
\label{complexity}
\begin{tabular}{ccc}
\toprule
  \makebox[0.12\textwidth][c]{Method} &\makebox[0.12\textwidth][c]{FLOPs} &\makebox[0.12\textwidth][c]{Params} \\
\midrule
DSIC & 178.4G &91.5M \\
HESIC$+$ & 48.6G &50.6M \\
DispSIC & 43.5G &15.9M \\
\bottomrule
\end{tabular}
\end{center}
% \vspace{-1.5em}
\end{table}

\begin{table}
\begin{center}
\caption{The specific configuration of ablation study.}
\label{ab-config}
% \resizebox{80mm}{20mm}{
\begin{tabular}{ccccc}
\toprule
  &Disparity &Priors &Aligned Priors &PRN \\
\midrule
{Case 1} & & & &  \\
{Case 2} & \checkmark & & &  \\
{Case 3} & \checkmark& \checkmark& & \\
{Case 4} & \checkmark& & \checkmark& \\
{DispSIC} & \checkmark& & \checkmark& \checkmark \\
\bottomrule
\end{tabular}
\end{center}
% \vspace{-1.5em}
\end{table}

\textbf{Evaluation.}
To evaluate the rate-distortion performance, we compare our model DispSIC with the learned stereo image compression methods \cite{liu2019dsic,deng2021deep}, the SOTA learned single image compression methods \cite{cheng2020image,hu2020coarse}, and the traditional image and video compression codecs \cite{bpgurl,sullivan2012overview,VVC} on the KITTI and InStereo2K dataset. It is worth mentioning that when comparing with HEVC \cite{sullivan2012overview}, stereo images are fed into the encoder as video sequences. The rate is measured by bits per pixel (bpp) and the quality is measured by peak signal-to-noise ratio (PSNR) and multi-scale structural similarity (MS-SSIM). Note the bpp is the average of both images. To compare the coding efficiency of different methods, we draw the rate-distortion (RD) curves. In addition, we report the Bjøntegaard delta PSNR (BD-PSNR) \cite{bjontegaard2001calculation} and BD-rate results to better compare methods. The higher value of BD-PSNR and the lower value of BD-rate indicate better image compression performance.

\subsection{Compression Results of Proposed Methods}

\noindent\textbf{Quantitative results.} 
Figure ~\ref{psnr_ssim} shows the rate-distortion performance of different methods on the KITTI and InStereo2K datasets. The data points on the RD curves are collected from the official GitHub page of HESIC \cite{deng2021deep}. For the MS-SSIM curve, in addition to the original scale, we also draw the log-scale, defined as $-10 \log_{10}$(1-MS-SSIM). Compared with the state-of-art (SOTA) stereo image compression method HESIC+ \cite{deng2021deep}, the SOTA learned image compression method Cheng \cite{cheng2020image}, and the SOTA traditional image compression codec VVC-intra \cite{VVC},  our method achieves the best performance on both datasets, which demonstrates the effectiveness and generality of our method. It is worth mentioning that our method achieves greater gain on the KITTI dataset at high bitrates. This demonstrates that our disparity-based method is more effective in scenes with complex structures. Table ~\ref{bdbr} shows the BD-BR and BD-PSNR results with DSIC \cite{liu2019dsic} as the baseline, we can see that our method achieves the highest BD-PSNR and the lowest BD-rate.

\noindent\textbf{Qualitative results.} 
In Fig. ~\ref{qualitative_results}, we provide some reconstructed images of different methods on the InStereo2K dataset.  For a fair comparison, all images are compressed to similar bit rates. As we can see, our model achieves higher PSNR of reconstructed images at lower bitrates. Meanwhile, it can be seen that our method can generate more structural details than all other methods, such as the stripes on the tiger and the lines on the basketball. 
% Due to the text limitations, the qualitative results on the KITTI dataset are shown in our supplementary materials.

\noindent\textbf{Computational complexity.}
In Table ~\ref{complexity}, we compare the FLOPs and parameters of DispSIC with HESIC+ and DSIC. Specifically, the resolution of the input image is $256 \times 256$. It can be seen that the Flops of DispSIC are slightly smaller than HESIC+ and much smaller than DSIC, and the number of parameters is much smaller than both HESIC+ and DSIC. The dense warp scheme of DSIC and the autoregressive model of HESIC+ for generating the H matrix have high computational complexity. Instead, we use a lightweight stereo matching 
model to explore the mutual information between stereo images, which greatly reduces the computational complexity. All in all, our model DispSIC has better performance and less complexity.

% \begin{table*}[t]
% \begin{center}
% \caption{Rate allocation of DispSIC on the  KITTI and InStereo2K datasets.}
% \label{rate-allocation}
% % \resizebox{80mm}{20mm}{
% \begin{tabular}{c|cc|cc|cc|cc}
% \toprule
% \makebox[0.08\textwidth][c]{$Dataset$} 
% &\makebox[0.08\textwidth][c]{$BPP$} &\makebox[0.08\textwidth][c]{$PSNR$}
% &\makebox[0.08\textwidth][c]{$BPP_R$} &\makebox[0.08\textwidth][c]{$PSNR_R$}
% &\makebox[0.08\textwidth][c]{$BPP_L$} &\makebox[0.08\textwidth][c]{$PSNR_L$}
% &\makebox[0.08\textwidth][c]{$BPP_D$} &\makebox[0.08\textwidth][c]{$PSNR_{R \rightarrow L}$}\\

% \midrule
%   \multirow{4}{*}{$KITTI$}
%       &0.177	&26.45	&0.222	&27.30  &0.118	&25.60	&0.014	&21.170 \\
%       &0.320	&28.60	&0.380	&29.35  &0.242	&27.86	&0.018	&21.493 \\
%       &0.488	&30.19	&0.560	&30.97  &0.395	&29.41	&0.022	&21.552 \\
%       &0.701	&31.72	&0.775	&32.55  &0.598	&30.90	&0.028	&21.678 \\
% \hline
%  \multirow{4}{*}{$InStereo2K$}
%  &0.109	&34.000	&0.177	&34.280 &0.033	&33.720	&0.009	&23.254\\
%  &0.245	&35.936	&0.363	&36.126 &0.116	&35.747	&0.010	&23.670\\
%  &0.393	&36.959	&0.535	&37.111 &0.239	&36.807	&0.013	&23.564\\
%  &0.651	&38.065	&0.710	&38.112 &0.577	&38.018	&0.015	&22.151\\
% \bottomrule
% \end{tabular}
% \end{center}
% % \vspace{-1.5em}
% \end{table*}

\begin{table*}[t]
\begin{center}
\caption{Rate allocation of DispSIC on the  KITTI dataset.}
\label{rate-allocation}
% \resizebox{80mm}{20mm}{
\begin{tabular}{c|cc|cc|cc|cc}
\toprule
\makebox[0.08\textwidth][c]{$Dataset$} 
&\makebox[0.08\textwidth][c]{$BPP$} &\makebox[0.08\textwidth][c]{$PSNR$}
&\makebox[0.08\textwidth][c]{$BPP_R$} &\makebox[0.08\textwidth][c]{$PSNR_R$}
&\makebox[0.08\textwidth][c]{$BPP_L$} &\makebox[0.08\textwidth][c]{$PSNR_L$}
&\makebox[0.08\textwidth][c]{$BPP_D$} &\makebox[0.08\textwidth][c]{$PSNR_{R \rightarrow L}$}\\

\midrule
  \multirow{4}{*}{$KITTI$}
      &0.177	&26.45	&0.222	&27.30  &0.118	&25.60	&0.014	&21.170 \\
      &0.320	&28.60	&0.380	&29.35  &0.242	&27.86	&0.018	&21.493 \\
      &0.488	&30.19	&0.560	&30.97  &0.395	&29.41	&0.022	&21.552 \\
      &0.701	&31.72	&0.775	&32.55  &0.598	&30.90	&0.028	&21.678 \\
% \hline
%  \multirow{4}{*}{$InStereo2K$}
%  &0.109	&34.000	&0.177	&34.280 &0.033	&33.720	&0.009	&23.254\\
%  &0.245	&35.936	&0.363	&36.126 &0.116	&35.747	&0.010	&23.670\\
%  &0.393	&36.959	&0.535	&37.111 &0.239	&36.807	&0.013	&23.564\\
%  &0.651	&38.065	&0.710	&38.112 &0.577	&38.018	&0.015	&22.151\\
\bottomrule
\end{tabular}
\end{center}
% \vspace{-1.5em}
\end{table*}
\subsection{Analysis of Rate Allocation}

% In Table ~\ref{rate-allocation}, we count the rate allocation of our method at different bitrates in the RD-curves on the KITTI and InStereo2K datasets. Note the calculation of ${BPP}_L$, ${BPP}_R$, and ${BPP}_D$ is the bitstream size divided by the number of pixels in a single image, and ${PSNR}_{R{\rightarrow}L}$ denotes the PSNR of the original left image and the predicted left image.
In Table ~\ref{rate-allocation}, we count the rate allocation of our method at different bitrates in the RD-curves on the KITTI dataset. Note the calculation of ${BPP}_L$, ${BPP}_R$, and ${BPP}_D$ is the bitstream size divided by the number of pixels in a single image, and ${PSNR}_{R{\rightarrow}L}$ denotes the PSNR of the left disparity-based prediction.

As we can see, our network assigns the most bitrates  ($BPP_R$) to the right image. 
As to the left image, we compress the left residuals and the disparity map.
The bitrates for the left residuals ($BPP_L$) are less.
Moreover, the disparity map is assigned the least bitrates ($BPP_D$).
Nonetheless, it can provide a respectable disparity compensated prediction, which greatly reduces the left residuals.
% As Fig.~\ref{entropy_image} shows, only some texture details are encoded as residuals into the bitstream.
In conclusion, by end-to-end optimization, our network can trade off the transmission cost and gain of the disparity map and adaptively allocate bitrates to the three parts to achieve the optimal compression performance.

\subsection{Ablation Study}
\begin{figure}
  \centering
  \includegraphics[width=0.917\linewidth, trim = {295 0 310 0}, clip]{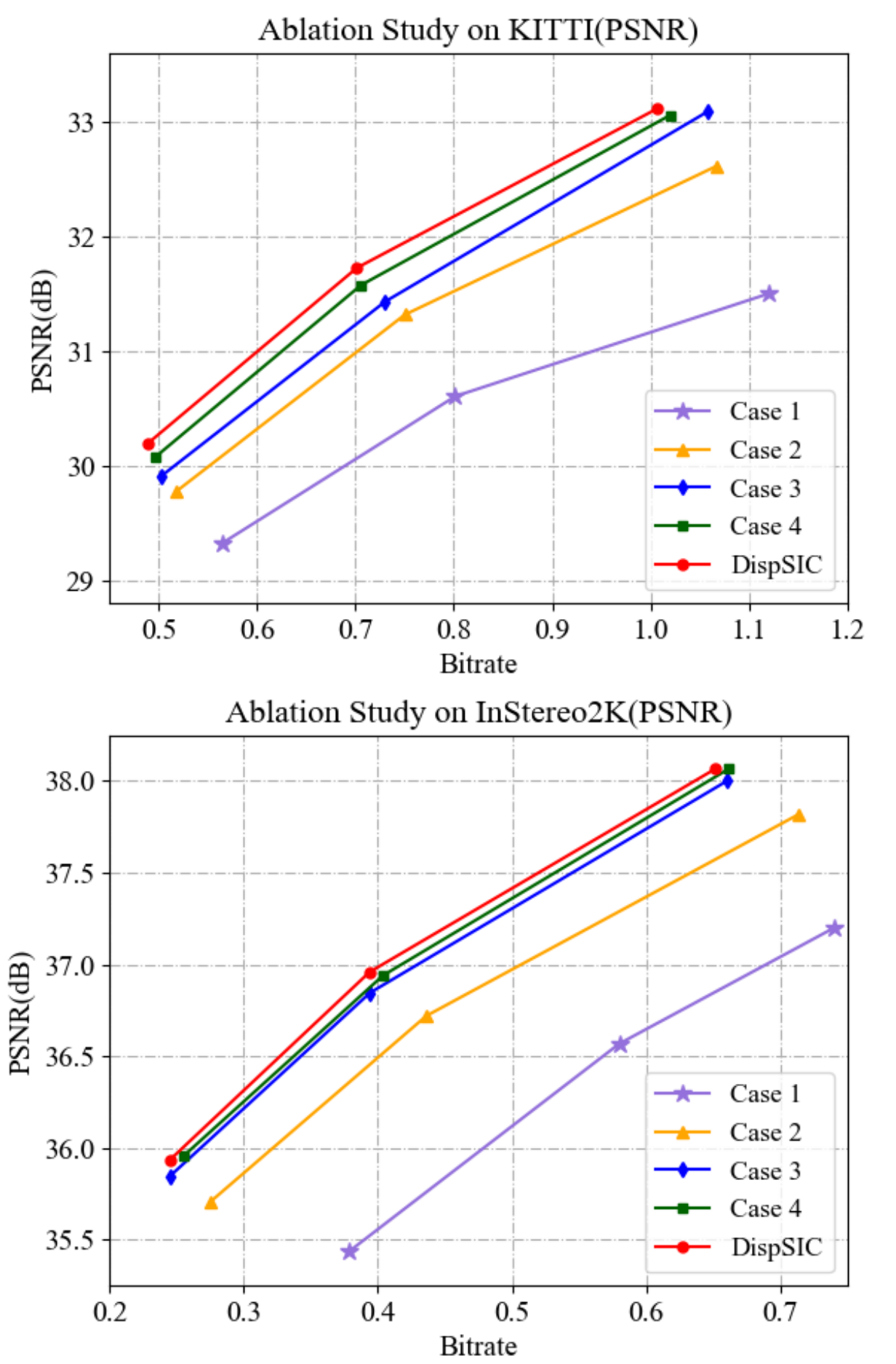}
  \caption{Ablation study of DispSIC on the KITTI and InStereo2K datasets. }
  \label{ablation}
\end{figure}

To analyze the contribution of each module, we implement ablation experiments on the KITTI and InStereo2K datasets as Figure ~\ref{ablation} shows.
Table ~\ref{ab-config} shows the specific configuration of ablation study.

\textbf{Disparity-based Prediction.} 
In case 1, we remove the disparity-based prediction and the priors. That means the stereo images are fed to the encoder separately, which can be considered as single image compression. As shown in Fig. ~\ref{ablation}, removing the disparity-based prediction causes significant performance degradation (case 1 vs. case 2), which demonstrates that the disparity-based prediction is the core module of our framework.

\textbf{Aligned Cross-View Priors.} 
To exploit the effectiveness of the aligned cross-view priors, we implement three experiments. In case 2, we simply remove the priors and use two independent entropy models for left and right images. In case 3, cross-view priors are sent to the left image entropy model, but lack alignment operation. In case 4, we align the cross-view priors with the left features.
As we can see, the RD performance has a large drop if priors are disabled (case 2 vs. case 3).
When enabling aligned cross-view priors (case 3 vs. case 4), the RD performance can be further improved, especially in scenes more complex.
These two comparative experiments demonstrate the effectiveness of our aligned cross-view priors.

\textbf{Prior Refinement Net.} 
Since the warping operation may introduce some spatial discontinuity, we design a prior refinement net to further improve the quality of the aligned cross-view priors. As shown in Fig. ~\ref{ablation}, the RD curve of case 4 is lower than the original model DispSIC. This result shows that the prior refinement net is helpful to improve the quality of the priors.
\section{Conclusion}
% In this paper, we propose a deep neural network for stereo image compression, which adopts the disparity to explicitly represent the pixel-wise relationship between the stereo image pair. Firstly, the disparity map is acquired through a stereo matching model, which warps the right image to the left view to get the residuals. Then the right image, the disparity map and the residuals of the left image are encoded into bitstream. We jointly train the stereo matching model and the compression model with the rate-distortion loss to achieve the optimal performance. 
In this paper, we propose a disparity-based stereo image compression network, which outperforms other existing SIC methods. 
% We jointly train a stereo matching model to assist in image compression. 
Based on the disparity map, the right image is warped to the left view to get the residual image.  Then the right image, the disparity map and the residual image are encoded into bitstream.
% Moreover, we propose a conditional entropy model with aligned cross-view priors, which warp the latent of right images to the left side.
% The warped latent is aligned with the left residuals' latent, which provides more consistent priors. Experiments on the KITTI and InStereo2K datasets show that our algorithm outperforms state-of-art learned stereo image compression methods.
Moreover, we propose a conditional entropy model with aligned cross-view priors, which provides more accurate probability estimation. Experiments on the KITTI and InStereo2K datasets show that our algorithm outperforms state-of-art learned stereo image compression methods.
\begin{acks}
This work is supported by National Natural Science Foundation of China U21B2012, 62072013 and 61902008, Shenzhen Cultivation of Excellent Scientific and Technological Innovation Talents RCJC20200714114435057, Shenzhen Research Projects of JCYJ2018\\0503182128089 and 201806080921419290. 
\end{acks}

\bibliographystyle{ACM-Reference-Format}
\bibliography{sample-sigconf}


\begin{thebibliography}{65}

%%% the Web address.

\ifx \showCODEN    \undefined \def \showCODEN     #1{\unskip}     \fi
\ifx \showDOI      \undefined \def \showDOI       #1{#1}\fi
\ifx \showISBNx    \undefined \def \showISBNx     #1{\unskip}     \fi
\ifx \showISBNxiii \undefined \def \showISBNxiii  #1{\unskip}     \fi
\ifx \showISSN     \undefined \def \showISSN      #1{\unskip}     \fi
\ifx \showLCCN     \undefined \def \showLCCN      #1{\unskip}     \fi
\ifx \shownote     \undefined \def \shownote      #1{#1}          \fi
\ifx \showarticletitle \undefined \def \showarticletitle #1{#1}   \fi
\ifx \showURL      \undefined \def \showURL       {\relax}        \fi
% invisible to TeX
\providecommand\bibfield[2]{#2}
\providecommand\bibinfo[2]{#2}
\providecommand\natexlab[1]{#1}
\providecommand\showeprint[2][]{arXiv:#2}

\bibitem[Agustsson et~al\mbox{.}(2017)]%
        {agustsson2017soft}
\bibfield{author}{\bibinfo{person}{Eirikur Agustsson}, \bibinfo{person}{Fabian
  Mentzer}, \bibinfo{person}{Michael Tschannen}, \bibinfo{person}{Lukas
  Cavigelli}, \bibinfo{person}{Radu Timofte}, \bibinfo{person}{Luca Benini},
  {and} \bibinfo{person}{Luc Van~Gool}.} \bibinfo{year}{2017}\natexlab{}.
\newblock \showarticletitle{Soft-to-hard vector quantization for end-to-end
  learning compressible representations}.
\newblock \bibinfo{journal}{\emph{arXiv preprint arXiv:1704.00648}}
  (\bibinfo{year}{2017}).
\newblock


\bibitem[Agustsson et~al\mbox{.}(2019)]%
        {agustsson2019generative}
\bibfield{author}{\bibinfo{person}{Eirikur Agustsson}, \bibinfo{person}{Michael
  Tschannen}, \bibinfo{person}{Fabian Mentzer}, \bibinfo{person}{Radu Timofte},
  {and} \bibinfo{person}{Luc~Van Gool}.} \bibinfo{year}{2019}\natexlab{}.
\newblock \showarticletitle{Generative adversarial networks for extreme learned
  image compression}. In \bibinfo{booktitle}{\emph{Proceedings of the IEEE/CVF
  International Conference on Computer Vision}}. \bibinfo{pages}{221--231}.
\newblock


\bibitem[Ball{\'e} et~al\mbox{.}(2017)]%
        {balle2017end-to-end}
\bibfield{author}{\bibinfo{person}{Johannes Ball{\'e}}, \bibinfo{person}{Valero
  Laparra}, {and} \bibinfo{person}{Eero~P Simoncelli}.}
  \bibinfo{year}{2017}\natexlab{}.
\newblock \showarticletitle{End-to-end Optimized Image Compression}.
\newblock  (\bibinfo{year}{2017}).
\newblock


\bibitem[Ball{\'e} et~al\mbox{.}(2018)]%
        {balle2018variational}
\bibfield{author}{\bibinfo{person}{Johannes Ball{\'e}}, \bibinfo{person}{David
  Minnen}, \bibinfo{person}{Saurabh Singh}, \bibinfo{person}{Sung~Jin Hwang},
  {and} \bibinfo{person}{Nick Johnston}.} \bibinfo{year}{2018}\natexlab{}.
\newblock \showarticletitle{Variational image compression with a scale
  hyperprior}.
\newblock \bibinfo{journal}{\emph{arXiv preprint arXiv:1802.01436}}
  (\bibinfo{year}{2018}).
\newblock


\bibitem[Bao et~al\mbox{.}(2020)]%
        {bao2020instereo2k}
\bibfield{author}{\bibinfo{person}{Wei Bao}, \bibinfo{person}{Wei Wang},
  \bibinfo{person}{Yuhua Xu}, \bibinfo{person}{Yulan Guo},
  \bibinfo{person}{Siyu Hong}, {and} \bibinfo{person}{Xiaohu Zhang}.}
  \bibinfo{year}{2020}\natexlab{}.
\newblock \showarticletitle{InStereo2K: A large real dataset for stereo
  matching in indoor scenes}.
\newblock \bibinfo{journal}{\emph{Science China Information Sciences}}
  \bibinfo{volume}{63}, \bibinfo{number}{11} (\bibinfo{year}{2020}),
  \bibinfo{pages}{1--11}.
\newblock


\bibitem[B{\'e}gaint et~al\mbox{.}(2020)]%
        {begaint2020compressai}
\bibfield{author}{\bibinfo{person}{Jean B{\'e}gaint}, \bibinfo{person}{Fabien
  Racap{\'e}}, \bibinfo{person}{Simon Feltman}, {and} \bibinfo{person}{Akshay
  Pushparaja}.} \bibinfo{year}{2020}\natexlab{}.
\newblock \showarticletitle{Compressai: a pytorch library and evaluation
  platform for end-to-end compression research}.
\newblock \bibinfo{journal}{\emph{arXiv preprint arXiv:2011.03029}}
  (\bibinfo{year}{2020}).
\newblock


\bibitem[Bezzine et~al\mbox{.}(2018)]%
        {bezzine2018sparse}
\bibfield{author}{\bibinfo{person}{I Bezzine}, \bibinfo{person}{Mounir
  Kaaniche}, \bibinfo{person}{Saadi Boudjit}, {and} \bibinfo{person}{Azeddine
  Beghdadi}.} \bibinfo{year}{2018}\natexlab{}.
\newblock \showarticletitle{Sparse optimization of non separable vector lifting
  scheme for stereo image coding}.
\newblock \bibinfo{journal}{\emph{Journal of Visual Communication and Image
  Representation}}  \bibinfo{volume}{57} (\bibinfo{year}{2018}),
  \bibinfo{pages}{283--293}.
\newblock


\bibitem[Birchfield and Tomasi(1999)]%
        {birchfield1999depth}
\bibfield{author}{\bibinfo{person}{Stan Birchfield} {and}
  \bibinfo{person}{Carlo Tomasi}.} \bibinfo{year}{1999}\natexlab{}.
\newblock \showarticletitle{Depth discontinuities by pixel-to-pixel stereo}.
\newblock \bibinfo{journal}{\emph{International Journal of Computer Vision}}
  \bibinfo{volume}{35}, \bibinfo{number}{3} (\bibinfo{year}{1999}),
  \bibinfo{pages}{269--293}.
\newblock


\bibitem[Bjontegaard(2001)]%
        {bjontegaard2001calculation}
\bibfield{author}{\bibinfo{person}{Gisle Bjontegaard}.}
  \bibinfo{year}{2001}\natexlab{}.
\newblock \showarticletitle{Calculation of average PSNR differences between
  RD-curves}.
\newblock \bibinfo{journal}{\emph{VCEG-M33}} (\bibinfo{year}{2001}).
\newblock


\bibitem[BPG(2019)]%
        {bpgurl}
\bibfield{author}{\bibinfo{person}{BPG}.} \bibinfo{year}{2019}\natexlab{}.
\newblock \bibinfo{title}{BPG Image format}.
\newblock \bibinfo{howpublished}{https://bellard.org/bpg/}.
\newblock


\bibitem[Chang and Chen(2018)]%
        {chang2018pyramid}
\bibfield{author}{\bibinfo{person}{Jia-Ren Chang} {and}
  \bibinfo{person}{Yong-Sheng Chen}.} \bibinfo{year}{2018}\natexlab{}.
\newblock \showarticletitle{Pyramid stereo matching network}. In
  \bibinfo{booktitle}{\emph{Proceedings of the IEEE conference on computer
  vision and pattern recognition}}. \bibinfo{pages}{5410--5418}.
\newblock


\bibitem[Cheng et~al\mbox{.}(2020)]%
        {cheng2020image}
\bibfield{author}{\bibinfo{person}{Zhengxue Cheng}, \bibinfo{person}{Heming
  Sun}, \bibinfo{person}{Masaru Takeuchi}, {and} \bibinfo{person}{Jiro Katto}.}
  \bibinfo{year}{2020}\natexlab{}.
\newblock \showarticletitle{Learned Image Compression with Discretized Gaussian
  Mixture Likelihoods and Attention Modules}. In
  \bibinfo{booktitle}{\emph{Proceedings of the IEEE Conference on Computer
  Vision and Pattern Recognition (CVPR)}}.
\newblock


\bibitem[Choi et~al\mbox{.}(2019)]%
        {choi2019variable}
\bibfield{author}{\bibinfo{person}{Yoojin Choi}, \bibinfo{person}{Mostafa
  El-Khamy}, {and} \bibinfo{person}{Jungwon Lee}.}
  \bibinfo{year}{2019}\natexlab{}.
\newblock \showarticletitle{Variable rate deep image compression with a
  conditional autoencoder}. In \bibinfo{booktitle}{\emph{Proceedings of the
  IEEE/CVF International Conference on Computer Vision}}.
  \bibinfo{pages}{3146--3154}.
\newblock


\bibitem[Christopoulos et~al\mbox{.}(2000)]%
        {2000The}
\bibfield{author}{\bibinfo{person}{Charilaos Christopoulos},
  \bibinfo{person}{A.~S. Skodras}, {and} \bibinfo{person}{Touradj Ebrahimi}.}
  \bibinfo{year}{2000}\natexlab{}.
\newblock \showarticletitle{The JPEG2000 still image coding system: An
  overview}.
\newblock \bibinfo{journal}{\emph{IEEE Transactions on Consumer Electronics}}
  \bibinfo{volume}{46}, \bibinfo{number}{4} (\bibinfo{year}{2000}),
  \bibinfo{pages}{1103--1127}.
\newblock


\bibitem[Cui et~al\mbox{.}(2021)]%
        {cui2021asymmetric}
\bibfield{author}{\bibinfo{person}{Ze Cui}, \bibinfo{person}{Jing Wang},
  \bibinfo{person}{Shangyin Gao}, \bibinfo{person}{Tiansheng Guo},
  \bibinfo{person}{Yihui Feng}, {and} \bibinfo{person}{Bo Bai}.}
  \bibinfo{year}{2021}\natexlab{}.
\newblock \showarticletitle{Asymmetric Gained Deep Image Compression With
  Continuous Rate Adaptation}. In \bibinfo{booktitle}{\emph{Proceedings of the
  IEEE/CVF Conference on Computer Vision and Pattern Recognition}}.
  \bibinfo{pages}{10532--10541}.
\newblock


\bibitem[Deng et~al\mbox{.}(2021)]%
        {deng2021deep}
\bibfield{author}{\bibinfo{person}{Xin Deng}, \bibinfo{person}{Wenzhe Yang},
  \bibinfo{person}{Ren Yang}, \bibinfo{person}{Mai Xu}, \bibinfo{person}{Enpeng
  Liu}, \bibinfo{person}{Qianhan Feng}, {and} \bibinfo{person}{Radu Timofte}.}
  \bibinfo{year}{2021}\natexlab{}.
\newblock \showarticletitle{Deep homography for efficient stereo image
  compression}. In \bibinfo{booktitle}{\emph{Proceedings of the IEEE/CVF
  Conference on Computer Vision and Pattern Recognition}}.
  \bibinfo{pages}{1492--1501}.
\newblock


\bibitem[Ellinas and Sangriotis(2004)]%
        {ellinas2004stereo}
\bibfield{author}{\bibinfo{person}{JN Ellinas} {and} \bibinfo{person}{Manolis~S
  Sangriotis}.} \bibinfo{year}{2004}\natexlab{}.
\newblock \showarticletitle{Stereo image compression using wavelet coefficients
  morphology}.
\newblock \bibinfo{journal}{\emph{Image and Vision Computing}}
  \bibinfo{volume}{22}, \bibinfo{number}{4} (\bibinfo{year}{2004}),
  \bibinfo{pages}{281--290}.
\newblock


\bibitem[Flierl and Girod(2007)]%
        {flierl2007multiview}
\bibfield{author}{\bibinfo{person}{Markus Flierl} {and} \bibinfo{person}{Bernd
  Girod}.} \bibinfo{year}{2007}\natexlab{}.
\newblock \showarticletitle{Multiview video compression}.
\newblock \bibinfo{journal}{\emph{IEEE signal processing magazine}}
  \bibinfo{volume}{24}, \bibinfo{number}{6} (\bibinfo{year}{2007}),
  \bibinfo{pages}{66--76}.
\newblock


\bibitem[Geiger et~al\mbox{.}(2013)]%
        {geiger2013vision}
\bibfield{author}{\bibinfo{person}{Andreas Geiger}, \bibinfo{person}{Philip
  Lenz}, \bibinfo{person}{Christoph Stiller}, {and} \bibinfo{person}{Raquel
  Urtasun}.} \bibinfo{year}{2013}\natexlab{}.
\newblock \showarticletitle{Vision meets robotics: The kitti dataset}.
\newblock \bibinfo{journal}{\emph{The International Journal of Robotics
  Research}} \bibinfo{volume}{32}, \bibinfo{number}{11} (\bibinfo{year}{2013}),
  \bibinfo{pages}{1231--1237}.
\newblock


\bibitem[Guo et~al\mbox{.}(2019)]%
        {guo2019group}
\bibfield{author}{\bibinfo{person}{Xiaoyang Guo}, \bibinfo{person}{Kai Yang},
  \bibinfo{person}{Wukui Yang}, \bibinfo{person}{Xiaogang Wang}, {and}
  \bibinfo{person}{Hongsheng Li}.} \bibinfo{year}{2019}\natexlab{}.
\newblock \showarticletitle{Group-wise correlation stereo network}. In
  \bibinfo{booktitle}{\emph{Proceedings of the IEEE/CVF Conference on Computer
  Vision and Pattern Recognition}}. \bibinfo{pages}{3273--3282}.
\newblock


\bibitem[He et~al\mbox{.}(2021)]%
        {he2021checkerboard}
\bibfield{author}{\bibinfo{person}{Dailan He}, \bibinfo{person}{Yaoyan Zheng},
  \bibinfo{person}{Baocheng Sun}, \bibinfo{person}{Yan Wang}, {and}
  \bibinfo{person}{Hongwei Qin}.} \bibinfo{year}{2021}\natexlab{}.
\newblock \showarticletitle{Checkerboard Context Model for Efficient Learned
  Image Compression}. In \bibinfo{booktitle}{\emph{Proceedings of the IEEE/CVF
  Conference on Computer Vision and Pattern Recognition}}.
  \bibinfo{pages}{14771--14780}.
\newblock


\bibitem[Hirschmuller(2007)]%
        {hirschmuller2007stereo}
\bibfield{author}{\bibinfo{person}{Heiko Hirschmuller}.}
  \bibinfo{year}{2007}\natexlab{}.
\newblock \showarticletitle{Stereo processing by semiglobal matching and mutual
  information}.
\newblock \bibinfo{journal}{\emph{IEEE Transactions on pattern analysis and
  machine intelligence}} \bibinfo{volume}{30}, \bibinfo{number}{2}
  (\bibinfo{year}{2007}), \bibinfo{pages}{328--341}.
\newblock


\bibitem[Hu et~al\mbox{.}(2020)]%
        {hu2020coarse}
\bibfield{author}{\bibinfo{person}{Yueyu Hu}, \bibinfo{person}{Wenhan Yang},
  {and} \bibinfo{person}{Jiaying Liu}.} \bibinfo{year}{2020}\natexlab{}.
\newblock \showarticletitle{Coarse-to-Fine Hyper-Prior Modeling for Learned
  Image Compression}. In \bibinfo{booktitle}{\emph{AAAI Conference on
  Artificial Intelligenc}}.
\newblock


\bibitem[Jaderberg et~al\mbox{.}(2015)]%
        {jaderberg2015spatial}
\bibfield{author}{\bibinfo{person}{Max Jaderberg}, \bibinfo{person}{Karen
  Simonyan}, \bibinfo{person}{Andrew Zisserman}, {et~al\mbox{.}}}
  \bibinfo{year}{2015}\natexlab{}.
\newblock \showarticletitle{Spatial transformer networks}.
\newblock \bibinfo{journal}{\emph{Advances in neural information processing
  systems}}  \bibinfo{volume}{28} (\bibinfo{year}{2015}).
\newblock


\bibitem[Jia et~al\mbox{.}(2019)]%
        {jia2019layered}
\bibfield{author}{\bibinfo{person}{Chuanmin Jia}, \bibinfo{person}{Zhaoyi Liu},
  \bibinfo{person}{Yao Wang}, \bibinfo{person}{Siwei Ma}, {and}
  \bibinfo{person}{Wen Gao}.} \bibinfo{year}{2019}\natexlab{}.
\newblock \showarticletitle{Layered image compression using scalable
  auto-encoder}. In \bibinfo{booktitle}{\emph{2019 IEEE Conference on
  Multimedia Information Processing and Retrieval (MIPR)}}. IEEE,
  \bibinfo{pages}{431--436}.
\newblock


\bibitem[Kadaikar et~al\mbox{.}(2018)]%
        {kadaikar2018joint}
\bibfield{author}{\bibinfo{person}{Aysha Kadaikar}, \bibinfo{person}{Gabriel
  Dauphin}, {and} \bibinfo{person}{Anissa Mokraoui}.}
  \bibinfo{year}{2018}\natexlab{}.
\newblock \showarticletitle{Joint disparity and variable size-block
  optimization algorithm for stereoscopic image compression}.
\newblock \bibinfo{journal}{\emph{Signal Processing: Image Communication}}
  \bibinfo{volume}{61} (\bibinfo{year}{2018}), \bibinfo{pages}{1--8}.
\newblock


\bibitem[Kendall et~al\mbox{.}(2017)]%
        {kendall2017end}
\bibfield{author}{\bibinfo{person}{Alex Kendall}, \bibinfo{person}{Hayk
  Martirosyan}, \bibinfo{person}{Saumitro Dasgupta}, \bibinfo{person}{Peter
  Henry}, \bibinfo{person}{Ryan Kennedy}, \bibinfo{person}{Abraham Bachrach},
  {and} \bibinfo{person}{Adam Bry}.} \bibinfo{year}{2017}\natexlab{}.
\newblock \showarticletitle{End-to-end learning of geometry and context for
  deep stereo regression}. In \bibinfo{booktitle}{\emph{Proceedings of the IEEE
  international conference on computer vision}}. \bibinfo{pages}{66--75}.
\newblock


\bibitem[Kingma and Ba(2014)]%
        {kingma2014adam}
\bibfield{author}{\bibinfo{person}{Diederik~P Kingma} {and}
  \bibinfo{person}{Jimmy Ba}.} \bibinfo{year}{2014}\natexlab{}.
\newblock \showarticletitle{Adam: A method for stochastic optimization}.
\newblock \bibinfo{journal}{\emph{arXiv preprint arXiv:1412.6980}}
  (\bibinfo{year}{2014}).
\newblock


\bibitem[Kitahara et~al\mbox{.}(2006)]%
        {kitahara2006multi}
\bibfield{author}{\bibinfo{person}{Masaki Kitahara}, \bibinfo{person}{Hideaki
  Kimata}, \bibinfo{person}{Shinya Shimizu}, \bibinfo{person}{Kazuto Kamikura},
  \bibinfo{person}{Yoshiyuki Yashima}, \bibinfo{person}{Kenji Yamamoto},
  \bibinfo{person}{Tomohiro Yendo}, \bibinfo{person}{Toshiaki Fujii}, {and}
  \bibinfo{person}{Masayuki Tanimoto}.} \bibinfo{year}{2006}\natexlab{}.
\newblock \showarticletitle{Multi-view video coding using view interpolation
  and reference picture selection}. In \bibinfo{booktitle}{\emph{2006 IEEE
  International Conference on Multimedia and Expo}}. IEEE,
  \bibinfo{pages}{97--100}.
\newblock


\bibitem[Klaus et~al\mbox{.}(2006)]%
        {klaus2006segment}
\bibfield{author}{\bibinfo{person}{Andreas Klaus}, \bibinfo{person}{Mario
  Sormann}, {and} \bibinfo{person}{Konrad Karner}.}
  \bibinfo{year}{2006}\natexlab{}.
\newblock \showarticletitle{Segment-based stereo matching using belief
  propagation and a self-adapting dissimilarity measure}. In
  \bibinfo{booktitle}{\emph{18th International Conference on Pattern
  Recognition (ICPR'06)}}, Vol.~\bibinfo{volume}{3}. IEEE,
  \bibinfo{pages}{15--18}.
\newblock


\bibitem[Kolmogorov and Zabih(2001)]%
        {kolmogorov2001computing}
\bibfield{author}{\bibinfo{person}{Vladimir Kolmogorov} {and}
  \bibinfo{person}{Ramin Zabih}.} \bibinfo{year}{2001}\natexlab{}.
\newblock \showarticletitle{Computing visual correspondence with occlusions
  using graph cuts}. In \bibinfo{booktitle}{\emph{Proceedings Eighth IEEE
  International Conference on Computer Vision. ICCV 2001}},
  Vol.~\bibinfo{volume}{2}. IEEE, \bibinfo{pages}{508--515}.
\newblock


\bibitem[Lee et~al\mbox{.}(2019)]%
        {lee2019context-adaptive}
\bibfield{author}{\bibinfo{person}{Jooyoung Lee}, \bibinfo{person}{Seunghyun
  Cho}, {and} \bibinfo{person}{Seungkwon Beack}.}
  \bibinfo{year}{2019}\natexlab{}.
\newblock \showarticletitle{Context-adaptive Entropy Model for End-to-end
  Optimized Image Compression}.
\newblock  (\bibinfo{year}{2019}).
\newblock


\bibitem[Liang et~al\mbox{.}(2018)]%
        {liang2018learning}
\bibfield{author}{\bibinfo{person}{Zhengfa Liang}, \bibinfo{person}{Yiliu
  Feng}, \bibinfo{person}{Yulan Guo}, \bibinfo{person}{Hengzhu Liu},
  \bibinfo{person}{Wei Chen}, \bibinfo{person}{Linbo Qiao}, \bibinfo{person}{Li
  Zhou}, {and} \bibinfo{person}{Jianfeng Zhang}.}
  \bibinfo{year}{2018}\natexlab{}.
\newblock \showarticletitle{Learning for disparity estimation through feature
  constancy}. In \bibinfo{booktitle}{\emph{Proceedings of the IEEE Conference
  on Computer Vision and Pattern Recognition}}. \bibinfo{pages}{2811--2820}.
\newblock


\bibitem[Liu et~al\mbox{.}(2019a)]%
        {liu2019non}
\bibfield{author}{\bibinfo{person}{Haojie Liu}, \bibinfo{person}{Tong Chen},
  \bibinfo{person}{Peiyao Guo}, \bibinfo{person}{Qiu Shen},
  \bibinfo{person}{Xun Cao}, \bibinfo{person}{Yao Wang}, {and}
  \bibinfo{person}{Zhan Ma}.} \bibinfo{year}{2019}\natexlab{a}.
\newblock \showarticletitle{Non-local attention optimized deep image
  compression}.
\newblock \bibinfo{journal}{\emph{arXiv preprint arXiv:1904.09757}}
  (\bibinfo{year}{2019}).
\newblock


\bibitem[Liu et~al\mbox{.}(2019b)]%
        {liu2019dsic}
\bibfield{author}{\bibinfo{person}{Jerry Liu}, \bibinfo{person}{Shenlong Wang},
  {and} \bibinfo{person}{Raquel Urtasun}.} \bibinfo{year}{2019}\natexlab{b}.
\newblock \showarticletitle{DSIC: Deep stereo image compression}. In
  \bibinfo{booktitle}{\emph{Proceedings of the IEEE/CVF International
  Conference on Computer Vision}}. \bibinfo{pages}{3136--3145}.
\newblock


\bibitem[Lukacs(1986)]%
        {lukacs1986predictive}
\bibfield{author}{\bibinfo{person}{M Lukacs}.} \bibinfo{year}{1986}\natexlab{}.
\newblock \showarticletitle{Predictive coding of multi-viewpoint image sets}.
  In \bibinfo{booktitle}{\emph{ICASSP'86. IEEE International Conference on
  Acoustics, Speech, and Signal Processing}}, Vol.~\bibinfo{volume}{11}. IEEE,
  \bibinfo{pages}{521--524}.
\newblock


\bibitem[Luo et~al\mbox{.}(2016)]%
        {luo2016efficient}
\bibfield{author}{\bibinfo{person}{Wenjie Luo}, \bibinfo{person}{Alexander~G
  Schwing}, {and} \bibinfo{person}{Raquel Urtasun}.}
  \bibinfo{year}{2016}\natexlab{}.
\newblock \showarticletitle{Efficient deep learning for stereo matching}. In
  \bibinfo{booktitle}{\emph{Proceedings of the IEEE conference on computer
  vision and pattern recognition}}. \bibinfo{pages}{5695--5703}.
\newblock


\bibitem[Ma et~al\mbox{.}(2021)]%
        {ma2021afec}
\bibfield{author}{\bibinfo{person}{Yi Ma}, \bibinfo{person}{Yongqi Zhai},
  \bibinfo{person}{Jiayu Yang}, \bibinfo{person}{Chunhui Yang}, {and}
  \bibinfo{person}{Ronggang Wang}.} \bibinfo{year}{2021}\natexlab{}.
\newblock \showarticletitle{AFEC: Adaptive Feature Extraction Modules for
  Learned Image Compression}. In \bibinfo{booktitle}{\emph{Proceedings of the
  29th ACM International Conference on Multimedia}}.
  \bibinfo{pages}{5436--5444}.
\newblock


\bibitem[Martinian et~al\mbox{.}(2006)]%
        {martinian2006view}
\bibfield{author}{\bibinfo{person}{Emin Martinian}, \bibinfo{person}{Alexander
  Behrens}, \bibinfo{person}{Jun Xin}, {and} \bibinfo{person}{Anthony Vetro}.}
  \bibinfo{year}{2006}\natexlab{}.
\newblock \showarticletitle{View synthesis for multiview video compression}. In
  \bibinfo{booktitle}{\emph{Picture Coding Symposium}},
  Vol.~\bibinfo{volume}{37}. \bibinfo{pages}{38--39}.
\newblock


\bibitem[Mayer et~al\mbox{.}(2016)]%
        {mayer2016large}
\bibfield{author}{\bibinfo{person}{Nikolaus Mayer}, \bibinfo{person}{Eddy Ilg},
  \bibinfo{person}{Philip Hausser}, \bibinfo{person}{Philipp Fischer},
  \bibinfo{person}{Daniel Cremers}, \bibinfo{person}{Alexey Dosovitskiy}, {and}
  \bibinfo{person}{Thomas Brox}.} \bibinfo{year}{2016}\natexlab{}.
\newblock \showarticletitle{A large dataset to train convolutional networks for
  disparity, optical flow, and scene flow estimation}. In
  \bibinfo{booktitle}{\emph{Proceedings of the IEEE conference on computer
  vision and pattern recognition}}. \bibinfo{pages}{4040--4048}.
\newblock


\bibitem[Mei et~al\mbox{.}(2021)]%
        {mei2021learning}
\bibfield{author}{\bibinfo{person}{Yixin Mei}, \bibinfo{person}{Li Li},
  \bibinfo{person}{Zhu Li}, {and} \bibinfo{person}{Fan Li}.}
  \bibinfo{year}{2021}\natexlab{}.
\newblock \showarticletitle{Learning-Based Scalable Image Compression with
  Latent-Feature Reuse and Prediction}.
\newblock \bibinfo{journal}{\emph{IEEE Transactions on Multimedia}}
  (\bibinfo{year}{2021}).
\newblock


\bibitem[Mentzer et~al\mbox{.}(2018)]%
        {mentzer2018conditional}
\bibfield{author}{\bibinfo{person}{Fabian Mentzer}, \bibinfo{person}{Eirikur
  Agustsson}, \bibinfo{person}{Michael Tschannen}, \bibinfo{person}{Radu
  Timofte}, {and} \bibinfo{person}{Luc Van~Gool}.}
  \bibinfo{year}{2018}\natexlab{}.
\newblock \showarticletitle{Conditional probability models for deep image
  compression}. In \bibinfo{booktitle}{\emph{Proceedings of the IEEE Conference
  on Computer Vision and Pattern Recognition (CVPR)}}.
  \bibinfo{pages}{4394--4402}.
\newblock


\bibitem[Mentzer et~al\mbox{.}(2020)]%
        {mentzer2020high}
\bibfield{author}{\bibinfo{person}{Fabian Mentzer}, \bibinfo{person}{George
  Toderici}, \bibinfo{person}{Michael Tschannen}, {and}
  \bibinfo{person}{Eirikur Agustsson}.} \bibinfo{year}{2020}\natexlab{}.
\newblock \showarticletitle{High-fidelity generative image compression}.
\newblock \bibinfo{journal}{\emph{arXiv preprint arXiv:2006.09965}}
  (\bibinfo{year}{2020}).
\newblock


\bibitem[Merkle et~al\mbox{.}(2006)]%
        {merkle2006efficient}
\bibfield{author}{\bibinfo{person}{Philipp Merkle}, \bibinfo{person}{Karsten
  Muller}, \bibinfo{person}{Aljoscha Smolic}, {and} \bibinfo{person}{Thomas
  Wiegand}.} \bibinfo{year}{2006}\natexlab{}.
\newblock \showarticletitle{Efficient compression of multi-view video
  exploiting inter-view dependencies based on H. 264/MPEG4-AVC}. In
  \bibinfo{booktitle}{\emph{2006 IEEE International Conference on Multimedia
  and Expo}}. IEEE, \bibinfo{pages}{1717--1720}.
\newblock


\bibitem[Minnen et~al\mbox{.}(2018)]%
        {minnen2018joint}
\bibfield{author}{\bibinfo{person}{David Minnen}, \bibinfo{person}{Johannes
  Ball{\'e}}, {and} \bibinfo{person}{George~D Toderici}.}
  \bibinfo{year}{2018}\natexlab{}.
\newblock \showarticletitle{Joint autoregressive and hierarchical priors for
  learned image compression}. In \bibinfo{booktitle}{\emph{Advances in Neural
  Information Processing Systems}}. \bibinfo{pages}{10771--10780}.
\newblock


\bibitem[Pilzer et~al\mbox{.}(2019)]%
        {pilzer2019progressive}
\bibfield{author}{\bibinfo{person}{Andrea Pilzer},
  \bibinfo{person}{St{\'e}phane Lathuili{\`e}re}, \bibinfo{person}{Dan Xu},
  \bibinfo{person}{Mihai~Marian Puscas}, \bibinfo{person}{Elisa Ricci}, {and}
  \bibinfo{person}{Nicu Sebe}.} \bibinfo{year}{2019}\natexlab{}.
\newblock \showarticletitle{Progressive fusion for unsupervised binocular depth
  estimation using cycled networks}.
\newblock \bibinfo{journal}{\emph{IEEE Transactions on Pattern Analysis and
  Machine Intelligence}} \bibinfo{volume}{42}, \bibinfo{number}{10}
  (\bibinfo{year}{2019}), \bibinfo{pages}{2380--2395}.
\newblock


\bibitem[Qian et~al\mbox{.}(2022)]%
        {qian2022entroformer}
\bibfield{author}{\bibinfo{person}{Yichen Qian}, \bibinfo{person}{Ming Lin},
  \bibinfo{person}{Xiuyu Sun}, \bibinfo{person}{Zhiyu Tan}, {and}
  \bibinfo{person}{Rong Jin}.} \bibinfo{year}{2022}\natexlab{}.
\newblock \showarticletitle{Entroformer: A Transformer-based Entropy Model for
  Learned Image Compression}.
\newblock \bibinfo{journal}{\emph{arXiv preprint arXiv:2202.05492}}
  (\bibinfo{year}{2022}).
\newblock


\bibitem[Rippel and Bourdev(2017)]%
        {rippel2017real}
\bibfield{author}{\bibinfo{person}{Oren Rippel} {and} \bibinfo{person}{Lubomir
  Bourdev}.} \bibinfo{year}{2017}\natexlab{}.
\newblock \showarticletitle{Real-time adaptive image compression}. In
  \bibinfo{booktitle}{\emph{International Conference on Machine Learning}}.
  PMLR, \bibinfo{pages}{2922--2930}.
\newblock


\bibitem[Ryan et~al\mbox{.}(1980)]%
        {ryan1980prediction}
\bibfield{author}{\bibinfo{person}{Thomas~W Ryan}, \bibinfo{person}{RT Gray},
  {and} \bibinfo{person}{Bobby~R Hunt}.} \bibinfo{year}{1980}\natexlab{}.
\newblock \showarticletitle{Prediction of correlation errors in stereo-pair
  images}.
\newblock \bibinfo{journal}{\emph{Optical Engineering}} \bibinfo{volume}{19},
  \bibinfo{number}{3} (\bibinfo{year}{1980}), \bibinfo{pages}{312--322}.
\newblock


\bibitem[Scharstein and Szeliski(2002)]%
        {scharstein2002taxonomy}
\bibfield{author}{\bibinfo{person}{Daniel Scharstein} {and}
  \bibinfo{person}{Richard Szeliski}.} \bibinfo{year}{2002}\natexlab{}.
\newblock \showarticletitle{A taxonomy and evaluation of dense two-frame stereo
  correspondence algorithms}.
\newblock \bibinfo{journal}{\emph{International journal of computer vision}}
  \bibinfo{volume}{47}, \bibinfo{number}{1} (\bibinfo{year}{2002}),
  \bibinfo{pages}{7--42}.
\newblock


\bibitem[Song et~al\mbox{.}(2020)]%
        {song2020edgestereo}
\bibfield{author}{\bibinfo{person}{Xiao Song}, \bibinfo{person}{Xu Zhao},
  \bibinfo{person}{Liangji Fang}, \bibinfo{person}{Hanwen Hu}, {and}
  \bibinfo{person}{Yizhou Yu}.} \bibinfo{year}{2020}\natexlab{}.
\newblock \showarticletitle{Edgestereo: An effective multi-task learning
  network for stereo matching and edge detection}.
\newblock \bibinfo{journal}{\emph{International Journal of Computer Vision}}
  \bibinfo{volume}{128}, \bibinfo{number}{4} (\bibinfo{year}{2020}),
  \bibinfo{pages}{910--930}.
\newblock


\bibitem[Su et~al\mbox{.}(2020)]%
        {su2020scalable}
\bibfield{author}{\bibinfo{person}{Rige Su}, \bibinfo{person}{Zhengxue Cheng},
  \bibinfo{person}{Heming Sun}, {and} \bibinfo{person}{Jiro Katto}.}
  \bibinfo{year}{2020}\natexlab{}.
\newblock \showarticletitle{Scalable Learned Image Compression With A Recurrent
  Neural Networks-Based Hyperprior}. In \bibinfo{booktitle}{\emph{2020 IEEE
  International Conference on Image Processing (ICIP)}}. IEEE,
  \bibinfo{pages}{3369--3373}.
\newblock


\bibitem[Sullivan and Ohm(2018)]%
        {VVC}
\bibfield{author}{\bibinfo{person}{G.~J. Sullivan} {and} \bibinfo{person}{J.~R.
  Ohm}.} \bibinfo{year}{2018}\natexlab{}.
\newblock \showarticletitle{{Versatile video coding Towards the next generation
  of video compression}}.
\newblock \bibinfo{journal}{\emph{Picture Coding Symposium}}
  (\bibinfo{year}{2018}).
\newblock


\bibitem[Sullivan et~al\mbox{.}(2012)]%
        {sullivan2012overview}
\bibfield{author}{\bibinfo{person}{Gary~J Sullivan},
  \bibinfo{person}{Jens-Rainer Ohm}, \bibinfo{person}{Woo-Jin Han}, {and}
  \bibinfo{person}{Thomas Wiegand}.} \bibinfo{year}{2012}\natexlab{}.
\newblock \showarticletitle{Overview of the high efficiency video coding (HEVC)
  standard}.
\newblock \bibinfo{journal}{\emph{IEEE Transactions on circuits and systems for
  video technology}} \bibinfo{volume}{22}, \bibinfo{number}{12}
  (\bibinfo{year}{2012}), \bibinfo{pages}{1649--1668}.
\newblock


\bibitem[Theis et~al\mbox{.}(2017)]%
        {theis2017lossy}
\bibfield{author}{\bibinfo{person}{Lucas Theis}, \bibinfo{person}{Wenzhe Shi},
  \bibinfo{person}{Andrew Cunningham}, {and} \bibinfo{person}{Ferenc
  Husz{\'a}r}.} \bibinfo{year}{2017}\natexlab{}.
\newblock \showarticletitle{Lossy image compression with compressive
  autoencoders}.
\newblock \bibinfo{journal}{\emph{arXiv preprint arXiv:1703.00395}}
  (\bibinfo{year}{2017}).
\newblock


\bibitem[Toderici et~al\mbox{.}(2015)]%
        {toderici2015variable}
\bibfield{author}{\bibinfo{person}{George Toderici}, \bibinfo{person}{Sean~M
  O'Malley}, \bibinfo{person}{Sung~Jin Hwang}, \bibinfo{person}{Damien
  Vincent}, \bibinfo{person}{David Minnen}, \bibinfo{person}{Shumeet Baluja},
  \bibinfo{person}{Michele Covell}, {and} \bibinfo{person}{Rahul Sukthankar}.}
  \bibinfo{year}{2015}\natexlab{}.
\newblock \showarticletitle{Variable rate image compression with recurrent
  neural networks}.
\newblock \bibinfo{journal}{\emph{arXiv preprint arXiv:1511.06085}}
  (\bibinfo{year}{2015}).
\newblock


\bibitem[Wallace(1992)]%
        {1992The}
\bibfield{author}{\bibinfo{person}{Gregory~K. Wallace}.}
  \bibinfo{year}{1992}\natexlab{}.
\newblock \showarticletitle{The JPEG still picture compression standard}.
\newblock \bibinfo{journal}{\emph{Communications of the Acm}}
  \bibinfo{volume}{38}, \bibinfo{number}{1} (\bibinfo{year}{1992}),
  \bibinfo{pages}{xviii--xxxiv}.
\newblock


\bibitem[Xie et~al\mbox{.}(2021)]%
        {xie2021enhanced}
\bibfield{author}{\bibinfo{person}{Yueqi Xie}, \bibinfo{person}{Ka~Leong
  Cheng}, {and} \bibinfo{person}{Qifeng Chen}.}
  \bibinfo{year}{2021}\natexlab{}.
\newblock \showarticletitle{Enhanced invertible encoding for learned image
  compression}. In \bibinfo{booktitle}{\emph{Proceedings of the 29th ACM
  International Conference on Multimedia}}. \bibinfo{pages}{162--170}.
\newblock


\bibitem[Xu and Zhang(2020)]%
        {xu2020aanet}
\bibfield{author}{\bibinfo{person}{Haofei Xu} {and} \bibinfo{person}{Juyong
  Zhang}.} \bibinfo{year}{2020}\natexlab{}.
\newblock \showarticletitle{Aanet: Adaptive aggregation network for efficient
  stereo matching}. In \bibinfo{booktitle}{\emph{Proceedings of the IEEE/CVF
  Conference on Computer Vision and Pattern Recognition}}.
  \bibinfo{pages}{1959--1968}.
\newblock


\bibitem[Yang et~al\mbox{.}(2018)]%
        {yang2018segstereo}
\bibfield{author}{\bibinfo{person}{Guorun Yang}, \bibinfo{person}{Hengshuang
  Zhao}, \bibinfo{person}{Jianping Shi}, \bibinfo{person}{Zhidong Deng}, {and}
  \bibinfo{person}{Jiaya Jia}.} \bibinfo{year}{2018}\natexlab{}.
\newblock \showarticletitle{Segstereo: Exploiting semantic information for
  disparity estimation}. In \bibinfo{booktitle}{\emph{Proceedings of the
  European conference on computer vision (ECCV)}}. \bibinfo{pages}{636--651}.
\newblock


\bibitem[Yu et~al\mbox{.}(2018)]%
        {yu2018deep}
\bibfield{author}{\bibinfo{person}{Lidong Yu}, \bibinfo{person}{Yucheng Wang},
  \bibinfo{person}{Yuwei Wu}, {and} \bibinfo{person}{Yunde Jia}.}
  \bibinfo{year}{2018}\natexlab{}.
\newblock \showarticletitle{Deep stereo matching with explicit cost aggregation
  sub-architecture}. In \bibinfo{booktitle}{\emph{Proceedings of the AAAI
  Conference on Artificial Intelligence}}, Vol.~\bibinfo{volume}{32}.
\newblock


\bibitem[Zbontar et~al\mbox{.}(2016)]%
        {zbontar2016stereo}
\bibfield{author}{\bibinfo{person}{Jure Zbontar}, \bibinfo{person}{Yann LeCun},
  {et~al\mbox{.}}} \bibinfo{year}{2016}\natexlab{}.
\newblock \showarticletitle{Stereo matching by training a convolutional neural
  network to compare image patches.}
\newblock \bibinfo{journal}{\emph{J. Mach. Learn. Res.}} \bibinfo{volume}{17},
  \bibinfo{number}{1} (\bibinfo{year}{2016}), \bibinfo{pages}{2287--2318}.
\newblock


\bibitem[Zhang and Wu(2021)]%
        {zhang2021attention}
\bibfield{author}{\bibinfo{person}{Xi Zhang} {and} \bibinfo{person}{Xiaolin
  Wu}.} \bibinfo{year}{2021}\natexlab{}.
\newblock \showarticletitle{Attention-guided Image Compression by Deep
  Reconstruction of Compressive Sensed Saliency Skeleton}. In
  \bibinfo{booktitle}{\emph{Proceedings of the IEEE/CVF Conference on Computer
  Vision and Pattern Recognition}}. \bibinfo{pages}{13354--13364}.
\newblock


\bibitem[Zhong et~al\mbox{.}(2017)]%
        {zhong2017self}
\bibfield{author}{\bibinfo{person}{Yiran Zhong}, \bibinfo{person}{Yuchao Dai},
  {and} \bibinfo{person}{Hongdong Li}.} \bibinfo{year}{2017}\natexlab{}.
\newblock \showarticletitle{Self-supervised learning for stereo matching with
  self-improving ability}.
\newblock \bibinfo{journal}{\emph{arXiv preprint arXiv:1709.00930}}
  (\bibinfo{year}{2017}).
\newblock


\bibitem[Zhu et~al\mbox{.}(2022)]%
        {zhu2021transformer}
\bibfield{author}{\bibinfo{person}{Yinhao Zhu}, \bibinfo{person}{Yang Yang},
  {and} \bibinfo{person}{Taco Cohen}.} \bibinfo{year}{2022}\natexlab{}.
\newblock \showarticletitle{Transformer-based Transform Coding}. In
  \bibinfo{booktitle}{\emph{International Conference on Learning
  Representations}}.
\newblock


\end{thebibliography}


\end{document}